\documentclass{jfm}

\usepackage{graphicx,amsmath,amssymb}

\usepackage{epstopdf, epsfig}
\usepackage{lineno,natbib}
\usepackage{colortbl} 

\newcommand{\infd}{\text{d}}

\newcommand{\pdiff}[2]{\frac{\partial{#1}}{\partial{#2}}}
\newcommand{\Ldiff}[2]{\frac{\text{D}#1}{\text{D}#2}}

\title{\Large Langmuir turbulence in suspended kelp farms}

\author{Tong Bo\aff{1}
  \corresp{\email{tbo@atmos.ucla.edu}},
  James C. McWilliams\aff{1}, 
  Chao Yan\aff{2} 
 \and Marcelo Chamecki\aff{1}}

\affiliation{\aff{1}Department of Atmospheric and Oceanic Sciences, 
University of California Los Angeles, Los Angeles,
CA 90095-1565, USA. 
\aff{2}Institute of Urban Meteorology, China Meteorological Administration, Beijing 100089, China.}

\begin{document}

\maketitle

\begin{abstract}
This study investigates the influence of suspended kelp farms on 
ocean mixed layer hydrodynamics in the presence of currents and waves. 
We use the large eddy simulation method, where the wave effect is 
incorporated by solving the wave-averaged equations. 
Distinct Langmuir circulation patterns are generated within various 
suspended farm configurations, including horizontally uniform kelp 
blocks and spaced kelp rows. 
Intensified turbulence arises from the farm-generated Langmuir 
circulation, as opposed to the standard Langmuir turbulence observed 
without a farm. The creation of Langmuir circulation within the farm 
is attributed to two primary factors depending on farm configuration: 
(1) enhanced vertical shear due to kelp frond area density variability, 
and (2) enhanced lateral shear due to canopy discontinuity at 
lateral edges of spaced rows. Both enhanced vertical and lateral shear 
of streamwise velocity, representing the lateral and vertical vorticity 
components respectively, can be tilted into downstream vorticity 
to create Langmuir circulation. This vorticity tilting is driven by the 
Craik-Leibovich vortex force associated with the Stokes drift of surface 
gravity waves. 
In addition to the farm-generated Langmuir turbulence, canopy shear 
layer turbulence is created at the farm bottom edge due to drag 
discontinuity. 
The intensity of different types of turbulence depends on both kelp 
frond area density and the geometric configuration of the farm. 
The farm-generated turbulence has substantial consequences for 
nutrient supply and kelp growth. 
These findings also underscore the significance of the presence 
of obstacle structures in modifying ocean mixed layer characteristics. 
\end{abstract}

\section{Introduction}
\label{sec:intro}
Marine macroalgae, such as kelp, provide essential habitats, 
shelter, and food sources for a diverse range of marine species, 
with immense importance for biodiversity preservation 
and ecosystem health \citep[e.g.,][]{dayton1985,teagle2017}. 
The cultivation and harvest of macroalgae also has the potential to become a sustainable 
strategy for biofuel production, food supply, and carbon sequestration 
\citep{ghadiryanfar2016,ferdouse2018}. Given the constraints posed by 
the ecological carrying capacity of existing nearshore aquaculture, 
recent interest has thus arisen in expanding macroalgal farming offshore 
\citep{troell2009,yan2021,frieder2022}. These offshore macroalgal farms 
are usually attached to suspended structures near the ocean surface, 
typically within the ocean mixed layer (OML). 

An essential factor affecting the performance of suspended macroalgal 
farms is their interaction with the hydrodynamic processes in the OML 
\citep{yan2021,frieder2022}. Kelp exerts a drag force on the flow 
\citep[e.g.,][]{thom1971,jackson1997}, resulting in attenuation in 
current velocity and wave motions \citep{rosman2007,monismith2022}. 
Discontinuities in drag can also lead to development of shear layers 
and eddies at the boundaries of the canopy \citep{plew2011depth}, which 
may cause significant modifications in OML turbulence \citep{yan2021}. 
These altered hydrodynamic conditions due to the presence of kelp can 
determine nutrient availability, chemical transport, and salinity 
and temperature conditions in the farms, thereby affecting kelp growth. 

Moreover, the variability of farm configurations, e.g., farm geometry 
and orientation with respect to currents and waves, can introduce  
added complexity into the interaction between kelp farms 
and OML turbulence. In addition, kelp growth and harvesting can 
effectively alter the frond surface area density \citep{frieder2022}, 
consequently influencing the drag force and canopy flow profiles. 
Comprehensive understanding of the complex hydrodynamic processes in 
the OML with the presence of suspended farms is therefore crucial for 
optimally designing farm configurations and tactically managing 
harvesting practices. 

The investigation of suspended farm hydrodynamics, beyond its direct 
implications for farm performance, also contributes to our broader 
understanding of how obstacle structures modify the OML. 
Various obstacle structures located near the ocean surface boundary, e.g., 
aquatic vegetation, engineered offshore platforms, ships, buoys, and 
sea ice, have the potential to influence the hydrodynamic interactions 
among winds, waves, and currents. 
These modifications to OML hydrodynamics may consequently result in 
alterations to other OML characteristics, e.g., heat transport 
and salinity mixing. 

Suspended kelp farms, hydrodynamically classified as suspended 
canopies, share similarities with submerged canopies that are 
located on the bottom boundary \citep{plew2011depth,tseung2016}. 
The emergence of shear layer turbulence at the top of the canopy has 
received considerable attention in the context of submerged canopy 
flow \citep[e.g.,][]{finnigan2000,nepf2012}. Likewise, for a suspended 
canopy, a shear layer can develop at the bottom of the canopy, 
leading to generation of turbulence and the exchange of momentum and 
scalars between the canopy and the underlying flow \citep{plew2011depth}. 
Additionally, for suspended canopies of finite dimensions, an adjustment 
region typically develops within the canopy starting from the leading edge 
\citep{tseung2016}, where the flow adapts to the drag imposed by the canopy, 
similar to that of finite-length submerged canopies \citep{belcher2003,rominger2011}. 
Subsequent to this adjustment region, a fully developed canopy flow 
region emerges, followed by a wake region downstream from the canopy. 
While the suspended canopy in general behaves like an inverted 
submerged canopy \citep{plew2011depth}, distinct hydrodynamic conditions 
can arise, as a result of the different boundary conditions at the 
surface and the prevalent presence of surface gravity waves in the OML. 

A prominent turbulent process in the OML is the Langmuir circulation driven 
by wind and waves \citep[e.g.,][]{leibovich1983,mcwilliams1997,thorpe2004}. 
It is typically visible on the ocean surface as streaks of foam or debris, 
i.e., surface convergence lines, and are usually roughly parallel to 
the directions of wind and waves \citep{langmuir1938}. The generation 
of Langmuir circulation depends critically on the interaction between 
the sheared wind-driven current and the Stokes drift of surface gravity 
waves \citep[e.g.,][]{craik1976,craik1977,leibovich1977}. Vorticity 
tilting due to the vertically sheared Stokes drift can cause instability 
to generate coherent Langmuir cell structures aligned with the downwind 
direction, and this is known as the Craik–Leibovich (CL)~2 mechanism. 
The intensified turbulence associated with Langmuir circulation can 
enhance the vertical transport and mixing in the upper ocean 
\citep{mcwilliams1997,mcwilliams2000} and could even be important for 
producing and maintaining the uniform OML \citep{li1997,thorpe2004}. 

Nevertheless, the interplay between Langmuir turbulence and aquatic 
vegetation remains largely unexplored. The significance of Langmuir 
circulation extends beyond hydrodynamics, also playing a crucial role 
in biogeochemical transport in the OML and influencing the 
distribution of macroalgae \citep{evans1980,qiao2009,dierssen2009}. 
On the other hand, the presence of vegetation can modify wind-driven 
currents and waves in the OML, and is thus expected to affect the 
generation of Langmuir turbulence. \citet{yan2021} investigated the 
creation of attached Langmuir circulation in a suspended macroalgal 
farm with a specific configuration, where spaced rows of kelp are 
aligned parallel to the flow and waves. In addition to the generation 
of canopy shear layer turbulence below the farm, Langmuir-type turbulence 
was found to occur within the farm, with a stronger magnitude than the 
standard Langmmuir turbulence generated without a farm. 
These various types of farm-generated turbulence can significantly 
affect nutrient transport in the OML, potentially leading to feedback 
on farm performance. Therefore, a more comprehensive examination of 
the physical mechanisms behind turbulence generation by suspended farms, 
e.g., the interaction between the Stokes drift and farm-modulated ocean 
currents, is necessary for an improved understanding of OML hydrodynamics. 
The dependence of canopy flow properties on different farm 
configurations also merits further investigation, as a crucial 
aspect of offshore farm planning and nutrient management. 

In this study, we employ large eddy simulation (LES) to understand 
the influence of suspended kelp farms on OML hydrodynamics, with 
a particular focus on the mechanisms of turbulence generation. 
We also explore how varying farm configurations and frond density 
distributions affect these turbulence generation mechanisms. 
Section~\ref{sec:methods} describes the numerical method and the various 
farm configurations investigated in this study. 
Section~\ref{sec:results} presents the statistics 
of mean flow, secondary flow, and turbulence in the farm. 
In particular, we focus on three representative farm configurations 
to highlight the various flow patterns arising from different 
horizontal arrangements and vertical frond density profiles.
Section~\ref{sec:tke-budget} examines the energy budget to understand 
sources of farm-generated turbulence. 
Section~\ref{sec:vort-dynamics} investigates the vorticity dynamics 
in the farm to illustrate the generation mechanisms of Langmuir 
circulations. In §~\ref{sec:other-factors}, we explore other 
farm parameters that affect turbulence generation, including 
the effective frond area density, farm orientation, and farm length. 
Section~\ref{sec:conclusion} presents the conclusion.

\section{Methods}
\label{sec:methods}
\subsection{Model description}
We use LES to investigate turbulence generation associated with  
suspended kelp farms. LES is a widely used tool in 
studying OML turbulence and more detailed discussion can be found 
in \citet{chamecki2019}. The code used in this study has been validated 
against Langmuir turbulence simulations in \citet{mcwilliams1997} and 
applied to previous research in the macroalgal farm and boundary 
layer flow \citep{yan2021,yan2022}.  The present LES framework is based 
on the wave-averaged and grid-filtered equations for mass, momentum, 
and heat: 
\begin{linenomath*}
\begin{equation}
\label{eq:mass}
    \nabla\cdot\tilde{\boldsymbol u}=0,
\end{equation}
\end{linenomath*}
\begin{linenomath*}
\begin{eqnarray}
\label{eq:mom}
\pdiff{\tilde{\boldsymbol u}}{t} 
+ \tilde{\boldsymbol u}\cdot\nabla\tilde{\boldsymbol u} 
& = & -\nabla\Pi - f\boldsymbol{e}_z\times\left(\tilde{\boldsymbol u} 
+ {\boldsymbol u}_s-{\boldsymbol u}_g\right) 
+ {\boldsymbol u}_s\times\tilde{\boldsymbol \zeta} \nonumber \\ 
&& + \left(1-\frac{\tilde\rho}{\rho_0}\right)g\boldsymbol{e}_z 
- \nabla\cdot\boldsymbol{\tau}^d - \boldsymbol F_D,
\end{eqnarray}
\end{linenomath*}
\begin{linenomath*}
\begin{equation}
\label{eq:heat}
    \pdiff{\tilde{\theta}}{t}+\left(\tilde{\boldsymbol u} 
    + {\boldsymbol u}_s\right) \cdot\nabla\tilde{\theta} 
    = - \nabla\cdot\boldsymbol{\pi}_\theta.
\end{equation}
\end{linenomath*}
This mathematical model was initially introduced in \citet{mcwilliams1997} 
by extending the original CL equations \citep{craik1976}, 
with the effects of planetary rotation and advection of scalars 
by Stokes drift incorporated. 

The tilde in \eqref{eq:mass}, \eqref{eq:mom}, and \eqref{eq:heat} 
represent the grid-filtered variables. 
In the Cartesian coordinate system $\boldsymbol{x} = (x,y,z)$, 
the velocity vector is $\tilde{\boldsymbol u}=(\tilde u,\tilde v,\tilde w)$, 
i.e., the streamwise, lateral (cross-stream), and vertical components, 
respectively. In \eqref{eq:mom}, $\Pi$ is the modified pressure 
\citep[e.g.,][]{chamecki2019}, $f$ is the Coriolis frequency, 
$g$ is the gravitational acceleration, $\boldsymbol{e}_z$ is the 
unit vector in the vertical direction, and $\tilde{\boldsymbol \zeta} 
=\nabla\times\tilde{\boldsymbol u}$ is the filtered vorticity. 
Here $\boldsymbol u_s$ is the Stokes drift associated with surface 
gravity waves, and $\boldsymbol u_g$ is a geostrophic current 
that represents the effect of mesoscale ocean flows. 
The geostrophic current is driven by an external pressure 
gradient $f\boldsymbol{e}_z\times{\boldsymbol u}_g$. 
The term $\boldsymbol F_D$ represents the drag force imposed by the 
canopy onto the flow, and the detailed treatment of canopy drag will 
be described later. 

In \eqref{eq:mom}, $\tilde\rho$ is the filtered density, and 
$\rho_0$ is the reference density. We assume that variations in 
density are only caused by the potential temperature $\tilde\theta$ 
via an linear relationship $\rho=\rho_0[1-\alpha(\theta-\theta_0)]$, 
where $\alpha=2\times10^{-4}$~K$^{-1}$ is the thermal exapansion 
coefficient and $\theta_0$ is the reference potential temperature 
corresponding to $\rho_0$. The term $\boldsymbol{\tau}^d$ is the 
deviatoric part of the subgrid-scale (SGS) stress tensor 
$\boldsymbol{\tau} = \widetilde{\boldsymbol u\boldsymbol u} - 
\tilde{\boldsymbol u}\tilde{\boldsymbol u}$, and $\boldsymbol{\pi}_\theta 
=\widetilde{\boldsymbol u\theta}-\tilde{\boldsymbol u}\tilde{\theta}$ 
is the SGS heat flux. The SGS stress is modeled using the Lagrangian 
scale-dependent dynamic Smagorinsky model \citep{bou2005}. 
The SGS heat flux is modeled using an eddy diffusivity closure, 
with diffusivity obtained from SGS viscosity and a prescribed value 
of turbulent Prandtl number $Pr_t = 0.4$. 
Molecular viscosity and diffusivity are assumed to be negligible for 
high Reynolds number flows examined in this study. 

The surface waves are not explicitly resolved in the model, and 
the time-averaged influences of waves are incorporated by imposing 
the Stokes drift $\boldsymbol{u}_s$. We consider 
a simple case of monochromatic deep water wave propagating in 
$x$~direction, with amplitude $a_w$ and frequency $\omega=\sqrt{gk}$, 
where $k$ is the wave number. The Stokes drift velocity thus reduces 
to $\boldsymbol{u}_s=(u_s,0,0)$, and 
\begin{linenomath*}
\begin{equation}
    \label{eq:stokes}
    u_s = U_s\textrm{e}^{2kz}, 
\end{equation}
\end{linenomath*}
where $U_s=\omega k a_w^2$ is the Stokes drift velocity at the surface. 
The effects of waves on OML turbulence 
is represented by the CL vortex force ${\boldsymbol u}_s\times
\tilde{\boldsymbol \zeta} = (0,-u_s\tilde\zeta_z,u_s\tilde\zeta_y)$, 
i.e., the third term on the right side of \eqref{eq:mom}. 
While the presence of the canopy may additionally influence waves, 
Stokes drift, and the wave-current interaction 
\citep[e.g.,][]{rosman2013,luhar2010,luhar2013}, these effects are 
estimated to be reasonably small for the macroalgal farm simulations 
considered here and thus have been neglected \citep{yan2021}. 

The parameterization of the canopy drag force $F_D$ in 
\eqref{eq:mom} is expressed as \citep{shaw1992,pan2014,yan2021}
\begin{linenomath*}
\begin{equation}
    \label{eq:drag}
    \boldsymbol{F}_D = \frac{1}{2}C_D a \boldsymbol{P}\cdot 
    \left(|\tilde{\boldsymbol u}|\tilde{\boldsymbol u}\right).
\end{equation}
\end{linenomath*}
Here $C_D$ is the drag coefficient, and $|\tilde{\boldsymbol u}|$ 
is the magnitude of the filtered velocity. Henceforth, for 
simplicity we will drop the tilde symbols that denote grid-filtered 
variables. We use $C_D=0.0148$ based on the experimental study of 
\citet{utter1996}, and more detailed discussion on this choice of 
$C_D$ can be found in \citet{yan2021}.
In \eqref{eq:drag}, $a$ is the frond surface area density 
(or foliage area density, area per volume, m$^{-1}$). 
The frond surface area of macroalgae is obtained by 
conversion of the algal biomass \citep{frieder2022}. Rather than 
directly resolving the geometry of the macroalgae fronds and stipes, 
their overall drag on the flow is characterized through this 
quadratic formula in the model, with a representative frond area 
density $a$ for each grid cell. 
The coefficient tensor $\boldsymbol{P}$ stands for the projection 
of frond surface area into each direction, and in the present 
study we use $\boldsymbol{P}=\frac{1}{2}\boldsymbol{I}$, where 
$\boldsymbol{I}$ is the identity matrix \citep{yan2021}. 

In the realistic farm setup, macroalgae are attached to subsurface 
structures at the bottom of the canopy \citep{charrier2018,yan2021}, 
and buoyancy typically keeps them upright in the water column 
\citep[e.g.,][]{koehl1977}. We therefore assume macroalgae fronds 
and stipes to maintain an approximately fixed position in 
the flow, except for exhibiting small-amplitude oscillations passively 
following the wave orbital motion. 
This assumption is supported by the analyses in \citet{yan2021}, 
which examined a set of dimensionless parameters including the 
Cauchy number and buoyancy number. It is also consistent 
with the omission of the wave orbital velocity in \eqref{eq:drag}, 
neglecting the interaction between wave and canopy drag 
\citep{yan2021,mcwilliams2023}. A more detailed description of 
macroalgal farm configurations will be presented in §~\ref{sec:farm-config}.

The present LES framework resolves the flow and temperature fields 
using a grid structure with horizontally collocated points and 
a vertically staggered arrangement. A pseudospectral method is used 
in the horizontal direction, and vertical derivatives are discretized 
with a second-order central finite-difference method. Aliasing errors 
arising from the nonlinear terms are removed through padding on the 
basis of the 3/2 rule. The equations are advanced in time using the 
fully explicit second-order Adams-Bashforth scheme. 

\subsection{Simulation setup}
The present study aims to investigate the 
mechanisms of turbulence generation in the OML in the presence of 
suspended kelp farms, with a focus on examining various farm 
configurations and frond area density distributions. Therefore, 
instead of considering a range of oceanic factors like wind stress, 
waves, currents, and mixed layer depth, we only select one set of 
representative oceanic conditions. The simulation parameters 
used here are generally the same as those in \citet{mcwilliams1997}. 
The flow is driven by a constant wind stress $\tau_w=0.037$~N~m$^{-2}$ 
at the surface boundary, corresponding to a wind speed at 10-m 
height above the surface of $U_{10}=5$~m~s$^{-1}$ and 
a friction velocity of $u_*=0.0061$~m~s$^{-1}$. 
The deep water waves have an amplitude of $a_w=0.8$~m, and the 
wavelength is $\lambda=60$~m, corresponding to a wave period of 6.2~s. 
This leads to a surface Stokes velocity of $U_s=0.068$~m~s$^{-1}$ 
and a turbulent Langmuir number of $La_t=\sqrt{u_*/U_s}=0.3$. 

In addition to the wind-driven current, a geostrophic flow 
$\boldsymbol{u}_g=(u_g,0,0)$ is imposed in $x$-direction, i.e., 
same as the direction of wind and waves. The geostrophic flow is 
driven by an external pressure gradient $fu_g$ in $y$-direction, 
with a constant value of $u_g=0.2$~m~s$^{-1}$, assuming that 
variations of mesoscale flow are small within the temporal and 
spatial scales relevant to the present study. The Coriolis frequency 
$f=10^{-4}$~s$^{-1}$ corresponds to 45$^\circ$~N latitude. 
The initial mixed layer depth of the upstream inflow is 25~m, 
and a stably stratified layer is beneath it, with a uniform temperature 
gradient $\infd \theta/\infd z=0.01$~K~m$^{-1}$ (buoyancy frequency 
$N=0.0044$~s$^{-1}$). We assume no heat flux at the surface boundary. 

\begin{figure}
  \centerline{\includegraphics[width=1\textwidth]{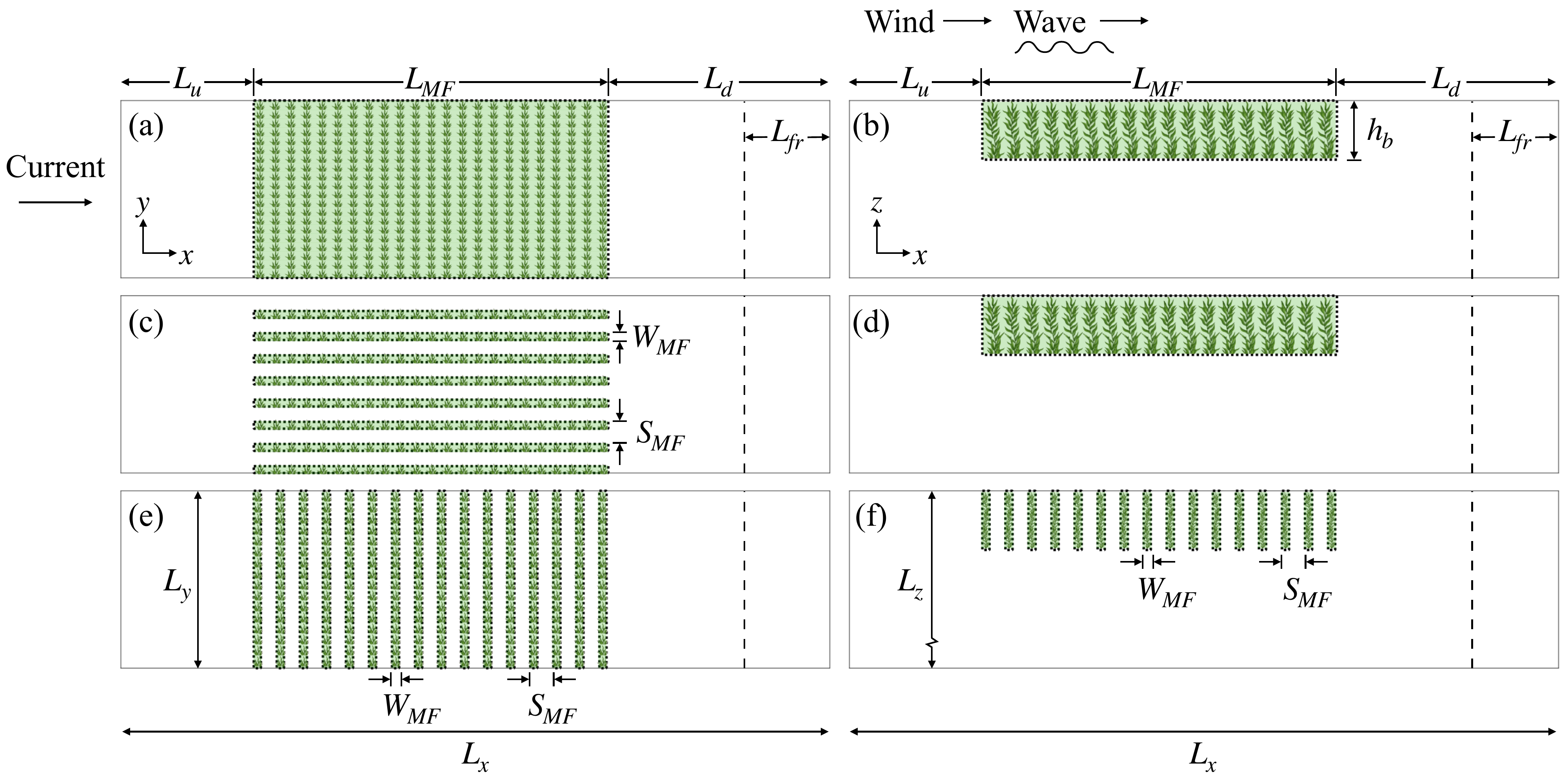}}
  \caption{Model domain and three types of farm configurations. 
  $(a)$ and $(b)$: The farm block configuration (top view and side view). 
  $(c)$ and $(d)$: The configuration with kelp rows aligned with the 
  current direction. $(e)$ and $(f)$: The configuration with kelp rows 
  oriented perpendicular to the current direction.}
\label{fig:model-domain}
\end{figure}

Kelp farm simulations are conducted on a $L_x\times L_y\times L_z=
800\times208\times120$~m$^3$ domain, with $N_x\times N_y\times N_z=
400\times104\times240$ grid cells. The mesh is uniformly distributed, 
with a horizontal resolution of 2~m and a vertical resolution of 0.5~m. 
A sensitivity test on grid size was conducted in \citet{yan2021}, 
and doubling the resolution in all the three dimensions yielded 
consistent results. 
The simulations are run for 15000~s to allow for the adjustment of 
the OML to the suspended canopy, and for another 9000~s after the 
fully developed turbulence state is reached to analyze turbulence 
statistics. A quasi-equilibrium state with converged turbulence 
statistics is established in the 9000~s period used for 
analysis, as examined in \citet{yan2021}. 

The farm is located in the middle of the domain from $x=0$ to $x=L_{MF}$, 
with a farm length of $L_{MF}=400$~m (figure~\ref{fig:model-domain}). 
The upstream boundary is at $x=-L_u=-150$~m, and the downstream 
boundary is at a distance of $L_d=250$~m from the farm trailing edge. 
In the $y$-direction the farm extends across the entire domain with a 
periodic boundary, i.e., effectively assuming an infinite farm width to 
eliminate the complexities arising from the lateral farm edges. 
In the vertical direction the farm is between the sea surface and 
$h_b=20$~m (the farm base), i.e., the depth at which the suspended 
structure is deployed. While most simulations are set as $L_x=800$~m 
and $L_{MF}=400$~m, an additional simulation is conducted with an 
extended domain length of $L_x=1200$~m (and $N_x=600$) and farm length 
of $L_{MF}=800$~m to examine the effect of a longer farm. 
More detailed descriptions of farm parameters will be presented 
in §~\ref{sec:farm-config}. 

A precursor inflow method is used to simulate the spatially evolving 
flow in the kelp farm simulations 
\citep{churchfield2012,stevens2014,yan2021}. In this method the 
turbulent velocity and temperature fields at the upstream boundary of 
the domain are obtained from a precursor simulation. The precursor 
simulation is separately conducted with identical conditions without 
the farm, until the turbulent flow reaches a quasi-equilibrium state. 
A fringe region (length $L_{fr}=100$~m) is used at the downstream end of 
the domain of the farm simulations. In this fringe region the flow field 
is smoothly forced toward the inflow conditions provided by the precursor 
simulation at the end of every time step. The precursor inflow method 
allows the turbulence produced by the precursor simulation to enter the 
domain of the farm simulations, while permitting the farm wakes to exit 
without cycling back through the periodic boundary conditions. 

\subsection{Farm configuration}
\label{sec:farm-config}
The cultivation of macroalgae in open ocean environments involves a 
diverse range of aquaculture structures. A representative farm 
configuration is considered here and consists of a series of organized 
longlines spaced horizontally \citep{yan2021,frieder2022}. 
Each longline is anchored at both ends and is also connected to surface 
buoys. Growth ropes, where kelp is seeded, are attached perpendicular 
to the longline. Kelp is cultivated at $h_b=20$~m, i.e., the longline 
deployment depth, and grows upright due to their buoyancy. The frond 
surface area density is assumed to be horizontally uniform within each 
canopy row (each longline set) for simplicity. In the present study, 
we explore various farm configurations by varying the canopy row 
spacing and orientation and by comparing different vertical 
profiles of frond surface area density within the row. 

\begin{figure}
  \centerline{\includegraphics[width=0.4\textwidth]{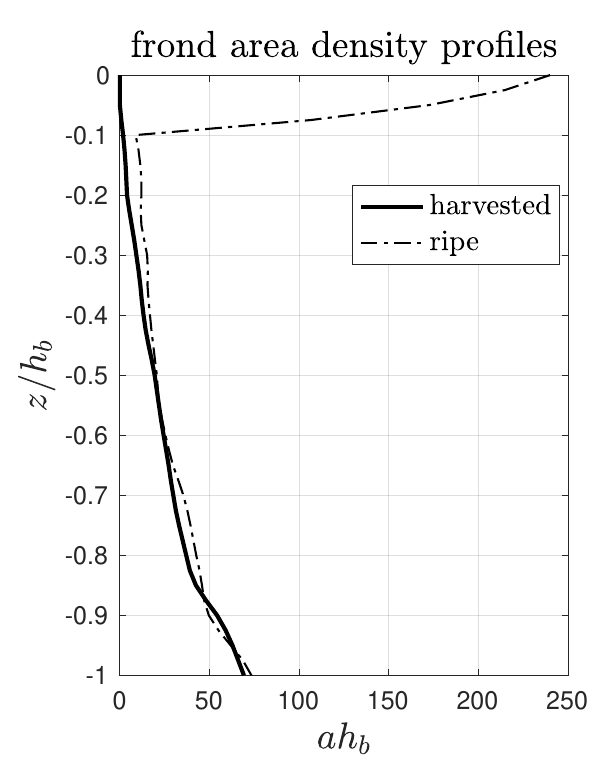}}
  \caption{Vertical profiles of frond surface area density $a$, 
  normalized by the farm base depth $h_b$. 
  The solid line represents the harvested profile, and the dash-dotted 
  line represents the ripe profile. The depth average value 
  $\langle a\rangle_z=1.14$~m$^{-1}$ ($\langle a\rangle_z h_b=23$) 
  for the harvested profile, and $\langle a\rangle_z=2.20$~m$^{-1}$ 
  ($\langle a\rangle_z h_b=44$) for the ripe profile.}
\label{fig:frond-prof}
\end{figure}

Two vertical profiles of frond surface area density are considered, which 
represent two different growth stages of kelp \citep{frieder2022}: 
(1) an intermediate growth stage with kelp extending from the farm base 
to around 2~m below the sea surface; 
(2) a fully grown stage with kelp extending from the farm base to the 
sea surface, with notably high frond area density at the top due to a 
large portion of the fronds floating at the sea surface. 
The frond surface area density profiles of the two stages are obtained 
by conversion of the algal biomass \citep{frieder2022} and are plotted 
in figure~\ref{fig:frond-prof}. 
Harvest practices typically concentrate on the uppermost 1-2~m 
part of kelp, resulting in the reduction of frond density to zero near 
the surface. This causes the frond density profile to revert to the 
earlier growth stage from the fully grown stage. Henceforth, we refer to 
the earlier growth stage -- with a low frond density near the surface -- 
as the `harvested' profile, and the fully grown stage -- with a high 
frond density near the surface -- is referred to as the `ripe' profile. 

In addition, we examine a variety of kelp row arrangements for each 
vertical profile of frond area density.  
The farm parameters for all the simulations are summarized in 
table~\ref{tab:para} in Appendix~\ref{app-farm-para}. 
The first set of arrangements (cases~`S') has 
the longlines aligned parallel to the $x$-direction (the direction of 
waves and geostrophic flow), extending the length of the farm. 
The longlines are repeated at a fixed distance $S_{MF}$ in the 
$y$-direction (figure~\ref{fig:model-domain}$c$ and $d$). 
We conduct a range of farm simulations with varying 
$S_{MF}$ and kelp row width (growth rope length) $W_{MF}$ 
(details provided in Appendix~\ref{app-farm-para}). 

The effective frond area density over the farm is defined as  
\begin{linenomath*}
\begin{equation}
\label{eq:eff-density}
    \left\langle a \right\rangle_{xyz} = \frac{1}{L_{MF}L_yh_b} \int_0^{L_{MF}}
    \int_{-L_y/2}^{L_y/2}\int_{-h_b}^{0} a\ \infd x\infd y\infd z.
\end{equation}
\end{linenomath*}
The frond area density $a$ takes the form depicted in 
figure~\ref{fig:frond-prof} within the kelp rows and is 0 in the 
gaps between rows. The effective frond density thus decreases 
as the spacing between kelp rows increases, while keeping the row 
width unchanged.  

Another set of farm arrangements (cases~`B') assumes a scenario where 
the kelp rows are positioned closely enough so that there is no 
gap in between, i.e., essentially forming a uniform kelp farm block 
(figure~\ref{fig:model-domain}$a$ and $b$). 
The two profiles in figure~\ref{fig:frond-prof} are employed in 
simulations involving this farm block configuration. 
Additionally, we conduct farm block simulations with varying 
effective density by introducing a multiplication factor to 
each frond density profile. 

In the third set of farm arrangements (cases~`PS'), the longlines 
are rotated by 90 degree, so that kelp rows are oriented parallel 
to the $y$-direction, perpendicular to the geostrophic flow 
(figure~\ref{fig:model-domain}$e$ and $f$). 
This set of arrangements also includes a range of values of 
$S_{MF}$ and $W_{MF}$, mirroring the above-mentioned simulations 
with longlines aligned parallel to the $x$-direction. 

Specifically, we have selected three simulations as representatives 
for an in-depth examination of the turbulence generation mechanisms. 
These selected cases are S26H (spaced farm rows parallel to the 
$x$-direction with a harvested profile, $S_{MF}=26$~m and $W_{MF}=8$~m; 
refer to table~\ref{tab:para}) and B1H and B1R (i.e., farm block 
simulations with harvested and ripe profiles, respectively). 
By comparing these three simulations, we aim to elucidate the 
impacts of cross-stream spacing and the vertical frond 
area density profile on OML turbulence generation. The effects of 
effective density and kelp row orientation will be discussed 
subsequently, following the analysis of turbulence generation 
mechanisms.

\section{Farm hydrodynamics}
\label{sec:results}
In this section we present the adjustment of mean flow to the 
kelp farm as well as the turbulence generated by the farm. 
We first introduce a flow decomposition to isolate distinct flow 
components. The instantaneous flow field can be split into the 
time-averaged and fluctuating components, i.e., 
\begin{linenomath*}
\begin{equation}
    \boldsymbol{u}(x,y,z,t) = \overline{\boldsymbol{u}}(x,y,z) + \boldsymbol{u}'(x,y,z,t).
\end{equation}
\end{linenomath*}
The overline represents the time average, and the prime represents temporal 
fluctuations about the time average. Further, the time-averaged flow 
field is decomposed into a cross-stream average and a steady 
cross-stream deviation, 
\begin{linenomath*}
\begin{equation}\label{eq:vel-decomp}
   \overline{\boldsymbol{u}}(x,y,z) = \left\langle\overline{\boldsymbol{u}}\right\rangle_y(x,z) 
   + \overline{\boldsymbol{u}}^c(x,y,z),
\end{equation}
\end{linenomath*}
where $\left\langle\cdot\right\rangle_y$ denotes the spatial averaging in 
$y$-direction. 

The temporal and cross-stream average 
$\left\langle\overline{\boldsymbol{u}}\right\rangle_y$ is defined as the 
mean flow. The steady cross-stream deviation $\overline{\boldsymbol{u}}^c$ 
is referred to as the secondary flow component, representing the 
stationary circulation structure generated by lateral variations in farm 
geometry. Note that $\overline{\boldsymbol{u}}^c$ only exists in 
farm configurations with laterally spaced farm rows and is negligible 
in the farm block or in kelp rows that are perpendicular to the 
geostrophic flow (details shown in following sections). 
The transient fluctuation $\boldsymbol{u}'$ represents the turbulence 
component. 

Similarly, the covariance between velocity and any field $\phi$ can 
be decomposed as 
\begin{linenomath*}
\begin{equation}\label{eq:flux-decomp}
   \left\langle\overline{\boldsymbol{u}\phi}\right\rangle_y = 
   \left\langle\overline{\boldsymbol{u}}\right\rangle_y\left\langle
   \overline{\phi}\right\rangle_y 
   + \left\langle\overline{\boldsymbol{u}}^c \overline{\phi}^c \right\rangle_y
   + \left\langle\overline{\boldsymbol{u}'\phi'}\right\rangle_y. 
\end{equation}
\end{linenomath*}
The first term on the right side stands for the contribution from 
the mean flow; the second term represents the effect of the secondary 
flow, akin to a dispersive flux \citep{finnigan2000}; the third term 
represents the turbulent flux. 

Specifically, we focus on three simulations, S26H, B1H, and B1R, as 
representatives to illustrate the influences of kelp row spacing in the 
cross-stream direction and the vertical profiles of frond area density.
Section~\ref{sec:results-mean} presents the adjustment of the time-mean 
flow field in the presence of kelp farms. 
Section~\ref{sec:overview-hydro} provides an overview of the hydrodynamic 
characteristics of all the different farm configurations. 
Section~\ref{sec:results-Langmuir} compares the Langmuir circulation 
patterns across the three representative cases. 
Subsequently, we quantify turbulence and steady secondary flows in the 
three cases in §~\ref{sec:results-energy}, and calculate 
the vertical velocity skewness in §~\ref{sec:results-skewness} to 
characterize the turbulence generated by the farm. 

\subsection{Mean flow}
\label{sec:results-mean}
\begin{figure}
  \centerline{\includegraphics[width=0.8\textwidth]{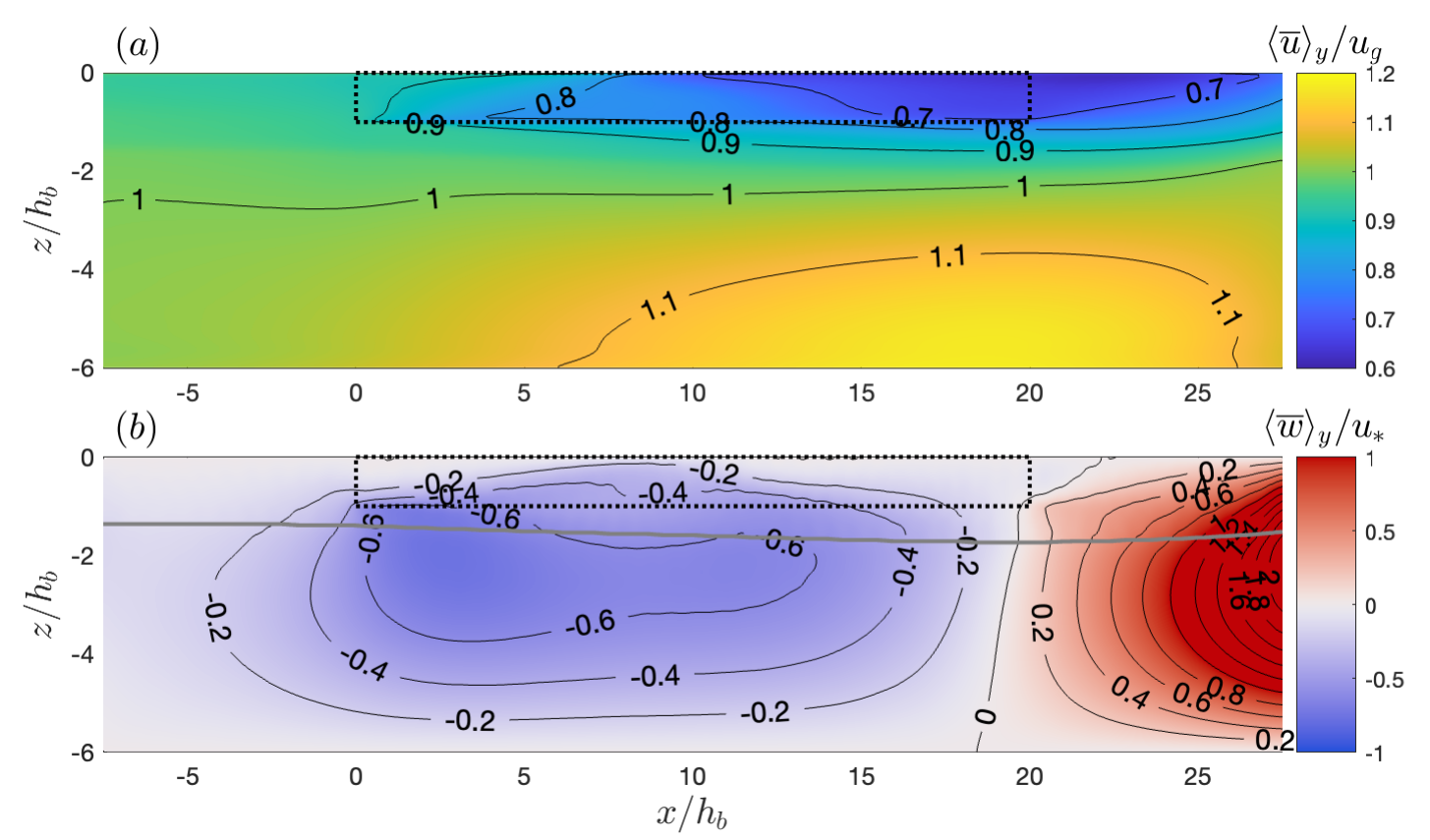}}
  \caption{Side views of mean flow in case S26H 
  (spaced rows aligned with the current, harvested profile). 
  $(a)$: Normalized streamwise velocity $\left\langle\overline{u}\right\rangle_y/u_g$. 
  $(b)$: Normalized vertical velocity $\left\langle\overline{w}\right\rangle_y/u_*$. 
  The mean flow is averaged in time and in the cross-stream direction. 
  Dotted rectangles show the extent of the farm, and the solid gray 
  line in $(b)$ represents the mixed layer depth. 
  Note that $\left\langle\overline{u}\right\rangle_y$ is normalized by $u_g$ 
  and $\left\langle\overline{w}\right\rangle_y$ is normalized by $u_*$, and 
  $\left\langle\overline{u}\right\rangle_y$ is generally much larger than 
  $\left\langle\overline{w}\right\rangle_y$.}
\label{fig:mean-vel-side}
\end{figure}

The mean flow structure is substantially altered by the presence 
of kelp farms. Figure~\ref{fig:mean-vel-side} shows the 
adjustment of mean flow in case S26H (spaced rows aligned with 
$x$-direction) as an example. As flow enters the canopy region, 
the streamwise velocity decreases due to the drag force exerted 
by the kelp (figure~\ref{fig:mean-vel-side}$a$). 
A shear layer develops beneath the farm as a result of 
discontinuity in kelp frond density. In addition, vertical shear is 
also increased within the canopy due to the vertical variability of 
frond area density (figure~\ref{fig:frond-prof}). The shear within 
the farm is most pronounced near the leading edge, and gradually 
diminishes downstream as turbulence generated in the farm leads to 
vertical mixing of momentum. 

At the canopy leading edge, as the streamwise flow is decelerated by 
the pressure gradient set up by the canopy drag, the mean downward 
vertical velocity develops as a result of mass conservation 
(figure~\ref{fig:mean-vel-side}$b$). 
Likewise, the pressure decrease at the trailing edge induces an 
upward velocity in the farm's wake region. Note that a fringe region 
is located between the wake region and the downstream end of the 
domain (not shown), where the flow field is forced toward the precursor 
inflow conditions to avoid the wake effect on periodic horizontal 
boundary conditions. 
The farm-induced mean flow adjustment in case S26H closely aligns with 
the previous study by \citet{yan2021}; other cases involving 
different farm configurations will be explored further. 

We define the mixed layer depth (MLD) $z_i$ as the depth at which the 
laterally averaged potential temperature 
$\left\langle\overline{\theta}\right\rangle_y$ first deviates from its 
surface value by $\Delta\theta$ \citep[e.g.,][]{kara2000}, here using 
$\Delta\theta = 0.01$~K. The MLD is around 25~m at the upstream 
boundary where the inflow originates. The mixed layer deepens to 
nearly 40~m due to the downwelling created by the farm, which then 
recovers downstream from the farm. The farm is always situated above 
the pycnocline, and the buoyancy effect is thus considered to have 
negligible influence on turbulence generation within the farm. 

\begin{figure}
  \centerline{\includegraphics[width=0.7\textwidth]{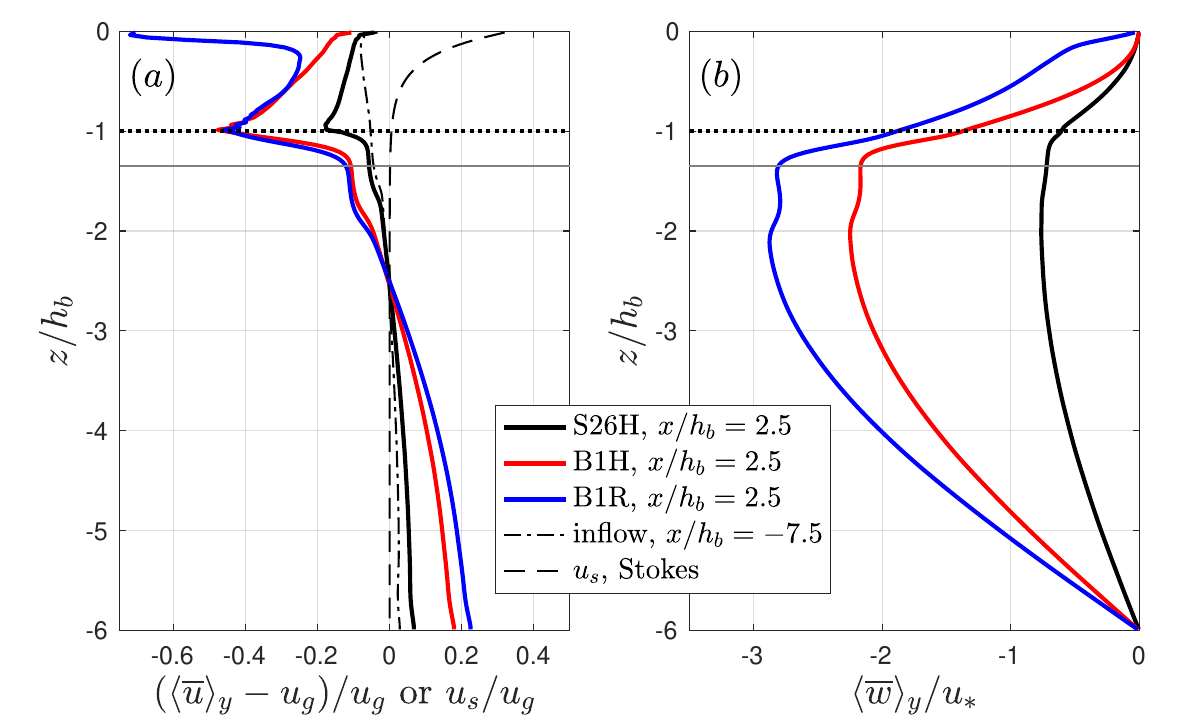}}
  \caption{Vertical profiles of streamwise velocity 
  $\left\langle\overline{u}\right\rangle_{y}$ $(a)$ and 
  vertical velocity $\left\langle\overline{w}\right\rangle_{y}$ $(b)$. 
  The velocities are time-averaged and cross-stream averaged, at $x/h_b=2.5$. 
  Note that the geostrophic current $u_g=0.2$~m/s has been subtracted in $(a)$. 
  The streamwise velocity in $(a)$ is normalized by $u_g$, and the vertical velocity 
  in $(b)$ is normalized by the friction velocity $u_*=0.0061$~m~s$^{-1}$.
  Black lines represent case S26H, spaced rows aligned with the 
  geostrophic current, with the harvested profile; red lines represent 
  case B1H, farm block with the harvested profile; blue lines represent 
  case B1R, farm block with the ripe profile. 
  Additionally, the dash-dotted line in $(a)$ shows the streamwise velocity 
  profile at the upstream boundary (inflow condition at $x/h_b=-7.5$). 
  The dashed line in $(a)$ represents the vertical profile of Stokes drift $u_s$. 
  The dotted horizontal lines mark the farm bottom, and the thin solid 
  horizontal lines represent the inflow mixed layer depth.}
\label{fig:mean-vel-prof}
\end{figure}

The canopy drag length $L_c$ is defined as 
\citep{belcher2003,rominger2011,belcher2012} 
\begin{linenomath*}
\begin{equation}
\label{eq:drag-length}
    L_c = \frac{2}{C_D \left\langle a\right\rangle_{xyz}},
\end{equation}
\end{linenomath*}
where $C_D$ is the drag coefficient, and $\left\langle a\right\rangle_{xyz}$ 
is the effective frond density in \eqref{eq:eff-density}. The factor 
of 2 in \eqref{eq:drag-length} accounts for the projection of frond 
surface area, as denoted by the term $\boldsymbol{P}$ in \eqref{eq:drag}. 
The canopy drag length is $L_c=347$~m for case S26H. Note that $L_c$ varies 
with the effective density and thus the farm configuration, and 
case B1R (farm block, ripe profile) has the smallest $L_c$ of 61~m. 
The length of the adjustment region scales in proportion to 
$L_c$, typically by a factor of $4.5-6$ \citep{belcher2012}. Therefore, 
the adjustment region length for the farm configurations in this study 
is either comparable to or larger than the farm length, suggesting 
that the canopy flow does not reach a fully developed state prior to 
exiting the farm. This explains the streamwise variations found 
in the mean flow field within the farm (figure~\ref{fig:mean-vel-side}) 
and also agrees with the analysis of the turbulence field presented 
in the following sections. 

The mean flow properties in the other simulations B1H and B1R are 
compared with S26H  (figure~\ref{fig:mean-vel-prof}). 
The vertical profiles in figure~\ref{fig:mean-vel-prof} are averaged 
in the cross-stream direction, and also averaged in the streamwise 
direction from the leading edge to $5h_b$ into the farm, 
a distance comparable to the canopy drag length $L_c$ in cases B1H 
and B1R. Generally cases B1H and B1R are similar to case S26H  
shown above, except that B1H and B1R demonstrate more pronounced downwelling 
at the leading edge and stronger vertical shear below the canopy, 
due to their higher effective frond area density and greater drag. 
In addition, the vertical shear of the mean streamwise flow within the farm 
is also stronger in B1H compared to S26H, because of the higher 
effective density in B1H. Moreover, while the streamwise mean velocity 
monotonically increases from the canopy bottom to the sea surface 
in cases S26H and B1H, the vertical shear reverses near the surface in B1R, 
because the presence of the uppermost dense layer in B1R locally 
enhances drag and decelerates flow. 

Note that the selected cases have a relatively high effective density 
$\left\langle a\right\rangle_{xyz}$. Other 
simulations that are not presented here, e.g., with larger spacing between 
kelp rows or with a lower frond area density, generally have weaker 
downwelling at the farm leading edge and weaker shear in the mean 
streamwsie flow. 

\subsection{Overview}
\label{sec:overview-hydro}
\begin{figure}
  \centerline{\includegraphics[width=1\textwidth]{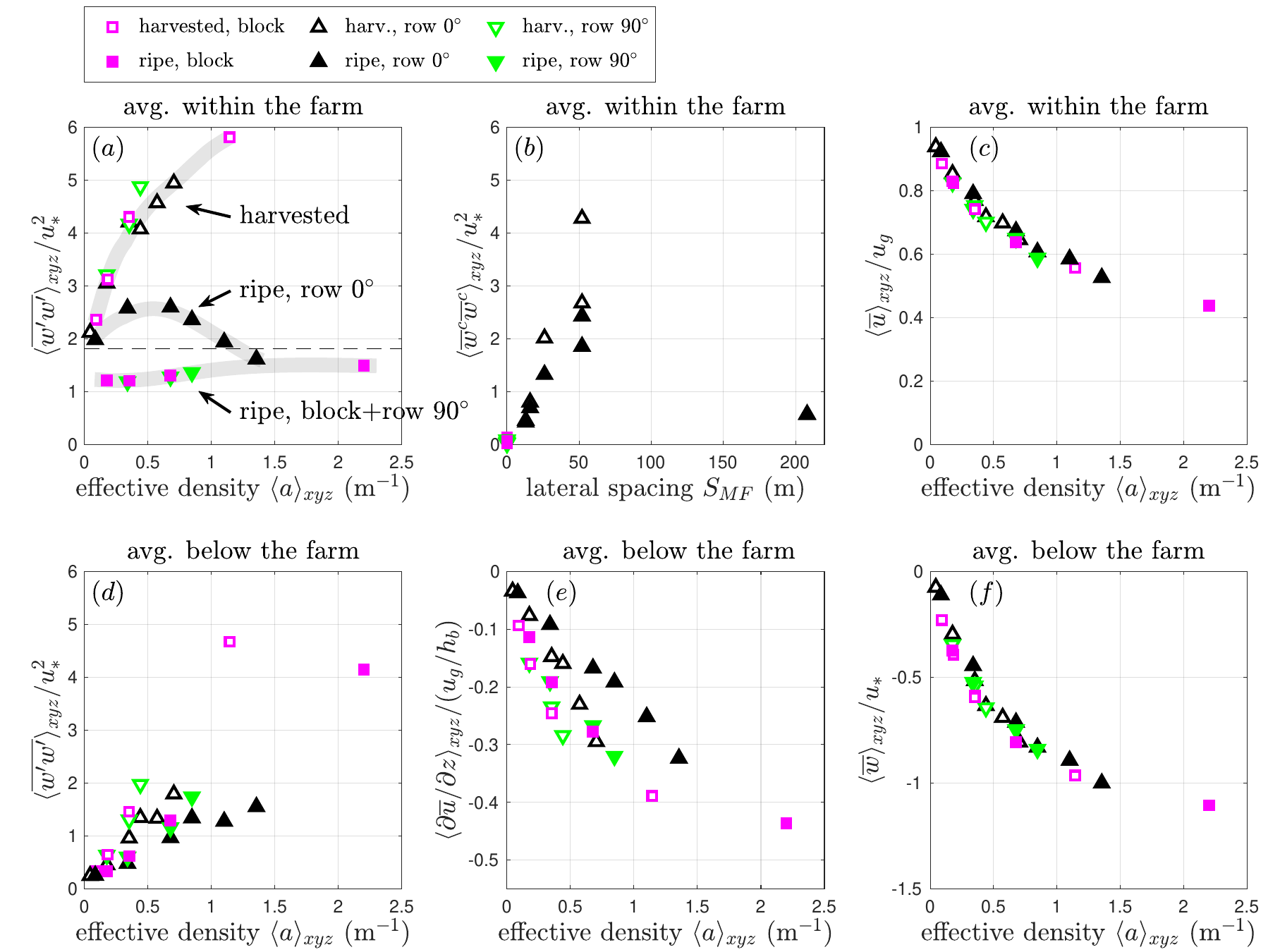}}
  \caption{Mean flow ($\overline u$, $\overline w$, and shear 
  $\partial\overline u/\partial z$), secondary flow ($\overline w^c$), 
  and turbulence ($w'$) statistics within the farm ($a$-$c$) and 
  below the farm ($d$-$f$). 
  Each point represents a simulation (see the legends for details). 
  Note that row~0$^\circ$ and row~90$^\circ$ denote kelp rows aligned with 
  and perpendicular to the current, respectively. 
  The horizontal dashed line in ($a$) represents the intensity of 
  standard Langmuir turbulence in the absence of a farm. 
  The thick gray lines in ($a$) are fitting curves for the three 
  types of farm configurations. 
  Averaging within the farm is conducted between $z=0$ and $-h_b$, 
  and averaging below the farm is between $-h_b$ and $-2h_b$. 
  The horizontal axis in ($a$, $c$-$f$) is the effective density 
  $\left\langle a \right\rangle_{xyz}$ averaged within the farm. 
  Note that the shear and vertical velocity are negative below the farm 
  in ($e$) and ($f$). The mean streamwise velocity is normalized by the 
  the geostrophic velocity $u_g=0.2$~m/s, the mean vertical velocity 
  is normalized by the friction velocity $u_*=0.0061$~m~s$^{-1}$, and 
  the variance terms are normalized by $u_*^2$.}
\label{fig:comp-tke}
\end{figure}
Along with modifications to the mean flow, the presence of the farm also 
affects Langmuir circulation patterns and turbulence intensity. 
In this section, we present an overview of the distinct flow patterns 
associated with various farm configurations, before moving on to 
detailed analysis of Langmuir circulation and turbulence. 
The magnitudes of mean flow, secondary flow, and turbulence components 
within and below the canopy are summarized in figure~\ref{fig:comp-tke} 
for all the farm configurations in table~\ref{tab:para}.

The mean streamwise flow within the farm decreases with the 
increased effective density $\left\langle a \right\rangle_{xyz}$ 
due to the greater kelp drag (figure~\ref{fig:comp-tke}$c$), 
leading to stronger downwelling below the farm as a result of 
mass conservation (figure~\ref{fig:comp-tke}$f$). 
In the shear layer below the canopy, the vertical shear of 
streamwise mean flow generally increases in magnitude with the 
increased effective frond density (figure~\ref{fig:comp-tke}$e$). 
Correspondingly, the turbulence intensity in the shear layer -- 
quantified by the variance of the temporal fluctuating component of 
vertical velocity $\left\langle \overline{w'w'} \right\rangle_{xyz}$ 
-- increases with the increased $\left\langle a \right\rangle_{xyz}$ 
(figure~\ref{fig:comp-tke}$d$). This is consistent with that expected 
for classical canopy flow, where the canopy effect positively depends on 
$\left\langle a \right\rangle_{xyz}$ \citep[e.g.,][]{poggi2004,bailey2013}. 
For shear layer statistics, the average is calculated within a vertical 
distance equal to $h_b$ beneath the canopy bottom, i.e., 
\begin{linenomath*}
\begin{equation}
    \left\langle \cdot \right\rangle_{xyz} = \frac{1}{L_{MF}L_yh_b} \int_0^{L_{MF}}
  \int_{-L_y/2}^{L_y/2}\int_{-2h_b}^{-h_b} \cdot \ \infd x\infd y\infd z 
\end{equation}
\end{linenomath*}
for averaging below the farm. 
Here we select the vertical component $w'$ because of its direct 
relevance to vertical transport in kelp farms. More discussions 
about the other components $u'$ and $v'$ and turbulence anisotropy 
will be presented later. 
Note that the scatter in figure~\ref{fig:comp-tke}$(d)$ and $(e)$ 
primarily results from the two different vertical profiles of frond area 
density. The dense layer near the surface in the ripe profile has a small 
influence on the shear layer dynamics below the canopy, and excluding 
this dense surface layer from the effective density calculation can lead 
to improved alignment in figure~\ref{fig:comp-tke}$(d)$ and $(e)$. 

As a contrast to the shear layer turbulence generated below the canopy, 
turbulence intensity within the farm displays a more complex dependence 
on the effective frond density (figure~\ref{fig:comp-tke}$a$). 
Note that within this depth range, e.g., between the sea surface 
and $z=-h_b$, standard Langmuir turbulence is expected to occur in the 
absence of the canopy \citep{mcwilliams1997}.
In the presence of the canopy, for kelp frond density with the harvested 
vertical profile, $\left\langle \overline{w'w'} \right\rangle_{y}$ is 
enhanced compared to that of the standard Langmuir turbulence. 
Moreover, the intensity of enhanced turbulence within the farm positively 
depends on the effective density $\left\langle a \right\rangle_{xyz}$. 
However, for farm configurations with the ripe vertical profile, 
turbulence is inhibited in farm blocks and kelp rows that are 
perpendicular to the geostrophic current. For farms with kelp rows aligned 
with the current, $\left\langle \overline{w'w'}\right\rangle_{y}$ is 
enhanced compared to the standard Langmuir turbulence only for cases with 
an intermediate effective density $\left\langle a \right\rangle_{xyz}$. 
This enhancement diminishes as $\left\langle a \right\rangle_{xyz}$ 
increases, i.e., with closely spaced rows that asymptotically 
resemble a farm block, or, as $\left\langle a \right\rangle_{xyz}$ 
decreases, asymptoting toward a scenario without a farm. 

For canopy flow without the influence of waves and Langmuir 
circulation, the turbulence intensity within the canopy is usually 
anticipated to decrease with increased effective density, due to the 
weaker penetration of shear layer eddies \citep{poggi2004,bailey2013}. 
Nevertheless, our simulations show that the turbulence intensity 
in the farm either positively depends on the effective density 
or exhibits a more complex dependence. 
This contrasting dependence on effective density underscores the 
distinct type of turbulence that arises from the interaction 
between the canopy and the OML, which is called the farm-generated 
Langmuir turbulence and will be presented in the following sections.

Moreover, secondary flow $\left\langle\overline{w}^c\overline{w}^c 
\right\rangle_{y}$ (stationary circulation) can be generated 
within the farm exclusively in cases with kelp rows aligned with 
the current (figure~\ref{fig:comp-tke}$b$). The occurrence of 
stationary secondary circulation associated with spaced kelp rows is 
consistent with the previous study by \citet{yan2021}, which referred to 
these flow patterns as `attached Langmuir circulation'. The intensity 
of $\left\langle\overline{w}^c\overline{w}^c \right\rangle_{y}$ is 
always stronger in cases with the harvested profile compared to 
the ripe profile. Our simulations do not allow for a clear 
identification of the maximum intensity with respect to the 
lateral spacing $S_{MF}$, and the peak of 
$\left\langle\overline{w}^c\overline{w}^c \right\rangle_{y}$ 
appears to fall within the range of $S_{MF}=52$~m and 208~m. 
Note that secondary flow never occurs in the shear layer below 
the canopy. 

Overall, these various flow statistics highlight the distinction 
between the hydrodynamics within the farm (within the OML where 
Langmuir turbulence is expected) and the classical shear layer 
turbulence beneath the farm. In addition, the contrast between 
different farm simulations suggests that these hydrodynamic processes 
depend on both the vertical frond density profile and horizontal farm 
arrangement, and these factors will be analyzed in detail below. 

\subsection{Langmuir circulations}
\label{sec:results-Langmuir}
\begin{figure}
  \centerline{\includegraphics[width=0.8\textwidth]{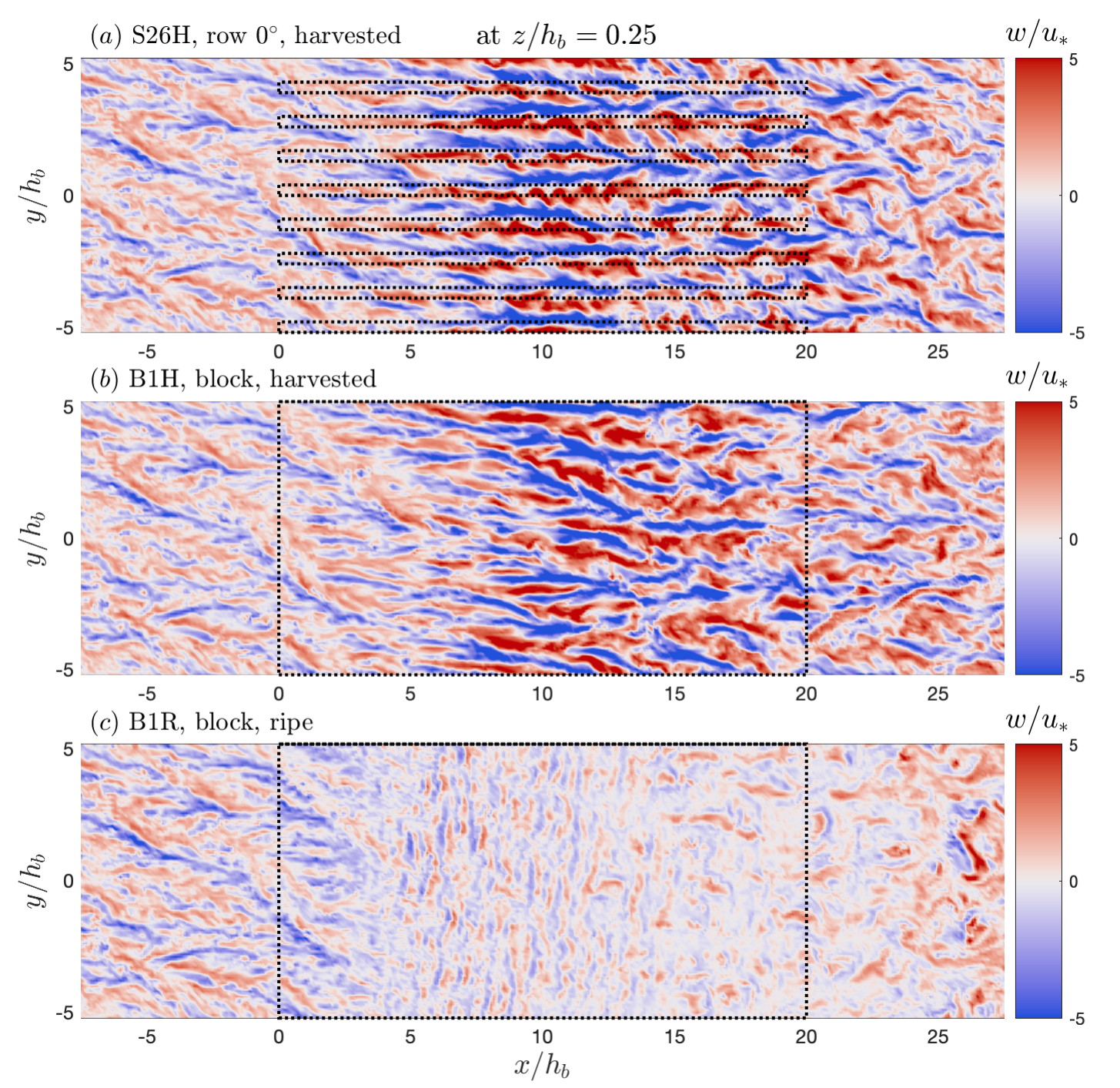}}
  \caption{Snapshots of normalized vertical velocity $w/u_*$ 
   on a horizontal plane at $z=-0.25h_b$, for cases S26H $(a)$, B1H $(b)$, 
   and B1R $(c)$. Dotted rectangles show the extent of the farm block or rows.}
\label{fig:map-w}
\end{figure}

Three representative simulations are selected to investigate the 
distinct flow patterns associated different farm configurations, 
i.e., case S26H (spaced rows aligned with $x$-direction, harvested 
profile), case B1H (farm block, harvested profile), and case B1R 
(farm block, ripe profile). Snapshots of instantaneous vertical 
velocity of the three simulations are compared in 
figure~\ref{fig:map-w}, on a horizontal plane at $z=-0.25h_b$. 

Upstream of the farm, the elongated streaks of downward vertical 
velocity are indicative of standard Langmuir circulation 
\citep[e.g.,][]{mcwilliams1997}. These patterns are typically 
featured by stronger downward motions within narrower regions 
compared to the broader and weaker upward motions. In addition, 
the streaks are generally rotated to the right of the direction of the 
wind and waves ($x$-direction) as a result of Coriolis. The standard 
Langmuir circulation patterns are transient, characterized by their 
continuous cycles of formation, evolution, and dissipation. 

For the case with laterally spaced kelp rows (case S26H), Langmuir 
circulation within the farm area shows a notable increase in magnitude 
compared to the standard Langmuir circulation in the upstream region 
(figure~\ref{fig:map-w}$a$). 
Moreover, these Langmuir circulation patterns are locked in space in 
the cross-stream direction, generally with upward motions within the 
kelp rows and downward motions within the gaps in between. 
The Langmuir patterns in the farm are generally stationary in time, 
although the smaller-scale coherent structures are still transient. 
These patterns are termed as the `attached Langmuir circulation' 
\citep{yan2021} because of their locked-in-space characteristics. 

For the farm block case with the harvested profile (case B1H), 
Langmuir circulation is also enhanced within the farm compared to 
the upstream region (figure~\ref{fig:map-w}$b$). However, these 
farm-enhanced Langmuir circulation patterns are completely transient 
and not locked in space due to the absence of repeated kelp rows, and 
we thus refer to these patterns as `unattached Langmuir circulation'. 
In the vertical direction, both the attached and unattached Langmuir 
circulation patterns have comparable dimentions to the farm height, 
with their maximum intensity found at a depth of around 10~m 
(quantitative results shown in §~\ref{sec:results-energy}). 

Furthermore, in contrast to the Langmuir patterns found in case B1H, 
Langmuir circulation notably vanishes in the farm in case B1R (farm block 
with the ripe profile, see figure~\ref{fig:map-w}$c$). 
This implies an absence of the Langmuir circulation generation mechanism 
or an increase in dissipation in case B1R, as will be investigated below. 
It is also worthwhile noting that the vertical velocity exhibits patterns 
aligned with the lateral direction in case B1R, in contrast to the 
streaks aligned with the streamwise direction in other cases. 
These distinct patterns in case B1R (e.g., for $x/h_b$ from 5 to 10) 
corresponds to shear-generated turbulence within the farm, as the canopy 
drag force significantly decelerates flow near the sea surface 
and enhances the vertical shear (figure~\ref{fig:mean-vel-prof}$a$). 
Moreover, shear layer turbulence generated at the farm bottom edge 
could penetrate into the farm in cases where Langmuir turbulence is 
inhibited (e.g., for $x/h_b>15$ in figure~\ref{fig:map-w}$c$). 
The distinct Langmuir patterns as well as the penetration of shear 
layer eddies will be further analyzed in subsequent sections. 

Additionally, a simulation with laterally spaced kelp rows with the ripe 
profile (case S26R) is examined. In this case, attached Langmuir circulation 
is generated in the farm (not shown) that resembles the patterns found in 
case S26H, while the magnitude of vertical velocity is smaller in case S26R 
compared to S26H. 

\subsection{Langmuir turbulence and bottom shear layer turbulence}
\label{sec:results-energy}
\begin{figure}
  \centerline{\includegraphics[width=0.8\textwidth]{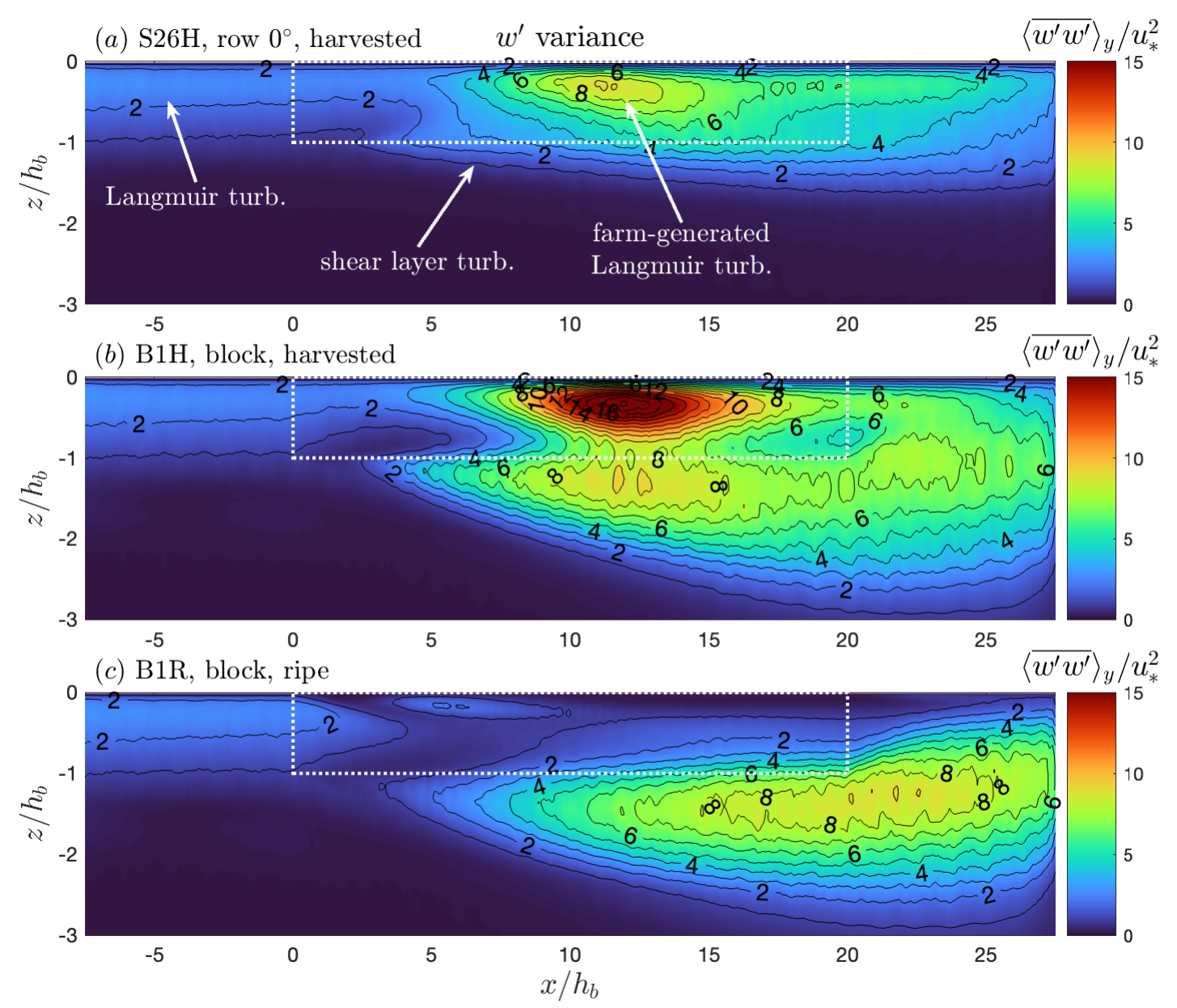}}
  \caption{Side views of the transient component of vertical velocity 
  variance $\left\langle\overline{w'w'}\right\rangle_{y}/u_*^2$ 
  for cases S26H $(a)$, B1H $(b)$, and B1R $(c)$. 
  The results are temporally and laterally averaged. 
  Dotted rectangles show the extent of the farm.}
\label{fig:sideview-wp2}
\end{figure}

The temporally and laterally averaged vertical velocity variance associated 
with transient eddies ($\left\langle\overline{w'w'}\right\rangle_{y}$) 
is calculated to quantify the farm-generated turbulence 
(figure~\ref{fig:sideview-wp2}). The vertical component $w'$ 
is selected because Langmuir turbulence is usually characterized by 
its large vertical velocity variance \citep{mcwilliams1997,yan2021}, 
and the vertical component also directly influences vertical transport. 

\begin{figure}
  \centerline{\includegraphics[width=0.7\textwidth]{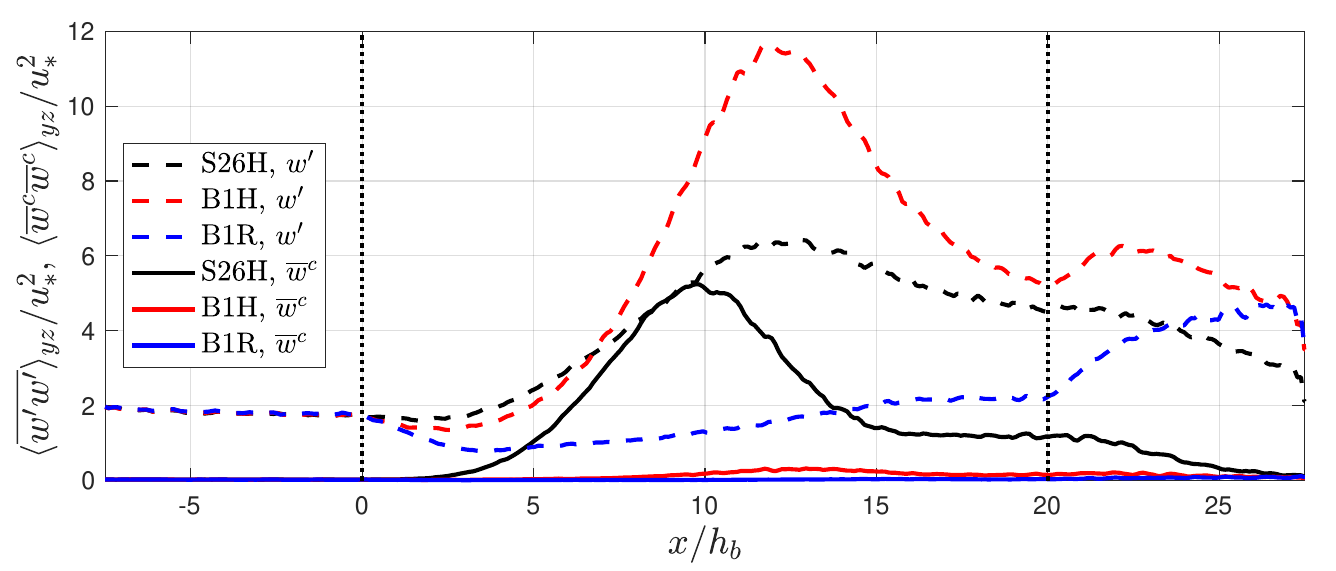}}
  \caption{Streamwise variations of 
  $\left\langle\overline{w'w'}\right\rangle_{y}/u_*^2$ (dashed lines)
  and $\left\langle\overline{w}^c \overline{w}^c\right\rangle_{y}/u_*^2$ (solid lines) 
  for cases S26H (black), B1H (red), and B1R (blue). 
  The results are temporally, laterally, and vertically ($z=0$ to $-h_b$) averaged. 
  Vertical dotted lines show the extent of the farm.}
\label{fig:along-wp2-wc2}
\end{figure}

Standard Langmuir turbulence occurs upstream of the farm, with the 
maximum $\left\langle\overline{w'w'}\right\rangle_{y}$ found slightly 
below the sea surface, e.g., around 5-10~m, consistent with 
\citet{mcwilliams1997}. Within the farm, similar Langmuir-type 
turbulence is generated in the two cases with the harvested profile 
(B1H and S26H, figure~\ref{fig:sideview-wp2}$a$ and $b$). The 
farm-generated Langmuir turbulence has a stronger magnitude compared 
to the standard Langmuir turbulence in the upstream region, while its 
vertical variance $\left\langle\overline{w'w'}\right\rangle_{y}$ 
also peaks at a similar depth of around 5-10~m. 
In the streamwise direction, the intensity of farm-generated Langmuir 
turbulence increases from the leading edge of the farm as flow adjusts 
to the canopy drag, with a maximum at around $x/h_b=10$ (also see 
figure~\ref{fig:along-wp2-wc2}). 
Turbulence intensity then decreases toward the farm trailing edge, 
due to the decrease in the production mechanisms and the damping by 
kelp drag and viscosity (details examined in the energy budget 
calculation in §~\ref{sec:tke-budget}). 
In addition, the farm-generated Langmuir turbulence is stronger in 
the farm block case B1H than the spaced rows case S26H (both with 
a harvested profile), and this aligns with the positive dependence 
on effective frond area density in figure~\ref{fig:comp-tke}$(a)$. 
By contrast, no intensified turbulence is found in the farm block 
with a ripe profile (case B1R) (figure~\ref{fig:sideview-wp2}$c$), 
consistent with the absence of Langmuir patterns in the map view plot 
of vertical velocity (figure~\ref{fig:map-w}$c$). 

The intensified turbulence in the kelp farm is referred to as 
Langmuir-type because its generation relates to the Stokes drift 
(see §~\ref{sec:tke-budget}), akin to the standard Langmuir 
turbulence. Furthermore, the intensified Langmuir-type turbulence in the farm 
completely disappears in another set of test cases (S26H-NW, B1H-NW, B1R-NW, 
not presented here), where the surface wave forcing is excluded. 
This provides corroborating evidence that the Langmuir-type turbulence 
results from the interaction between waves and canopy flow. 

In addition to the Langmuir-type turbulence generated within the farm, 
shear layer turbulence occurs at the bottom edge of the canopy. The 
shear layer turbulence, as a characteristic of classical canopy flow, 
is consistently found across all the simulations regardless of the 
presence or absence of Langmuir-type turbulence within the farm. 
The intensity of shear layer turbulence typically increases with the 
increased effective density \citep[e.g.,][]{bailey2013}, so that 
the farm block cases (B1H and B1R) exhibit stronger turbulence below 
the farm compared to the case with spaced kelp rows (S26H). 

While the shear layer turbulence mostly occurs beneath the canopy, 
it also penetrates into the canopy, in particular in the downstream 
part of the farm. Nevertheless, the penetrated shear layer turbulence 
is overall much weaker than the farm-generated Langmuir turbulence 
(as will be substantiated by the skewness analysis in 
§~\ref{sec:results-skewness}). Therefore, in cases with enhanced Langmuir 
turbulence, shear layer turbulence has a minimal influence on the turbulence 
statistics calculated within the farm. This point is also supported by 
the dependence of turbulence intensity on the effective density 
$\left\langle a \right\rangle_{xyz}$ in figure~\ref{fig:comp-tke}$(a)$. 
Turbulence intensity within the canopy is expected to decrease with 
increased $\left\langle a \right\rangle_{xyz}$, assuming penetration 
of shear layer is the dominant source of turbulence 
\citep{poggi2004,bailey2013}. Nevertheless, our simulations show a 
more complex dependence of turbulence intensity on 
$\left\langle a \right\rangle_{xyz}$, due to the dominance of 
Langmuir-type turbulence within the farm. 

We have primarily focused on the vertical component $w'$, and the 
horizontal components $u'$ and $v'$ generally display similar 
distributions to $w'$. However, it is worthwhile noting that the 
transient eddies are not isotropic, and this turbulence anisotropy 
also provides insights into distinguishing between different types 
of farm-generated turbulence. The shear layer turbulence typically 
has a stronger streamwise component $u'$  compared to the other 
components, while the Langmuir-type turbulence is featured by a 
relatively stronger vertical component $w'$. 

\begin{figure}
  \centerline{\includegraphics[width=0.8\textwidth]{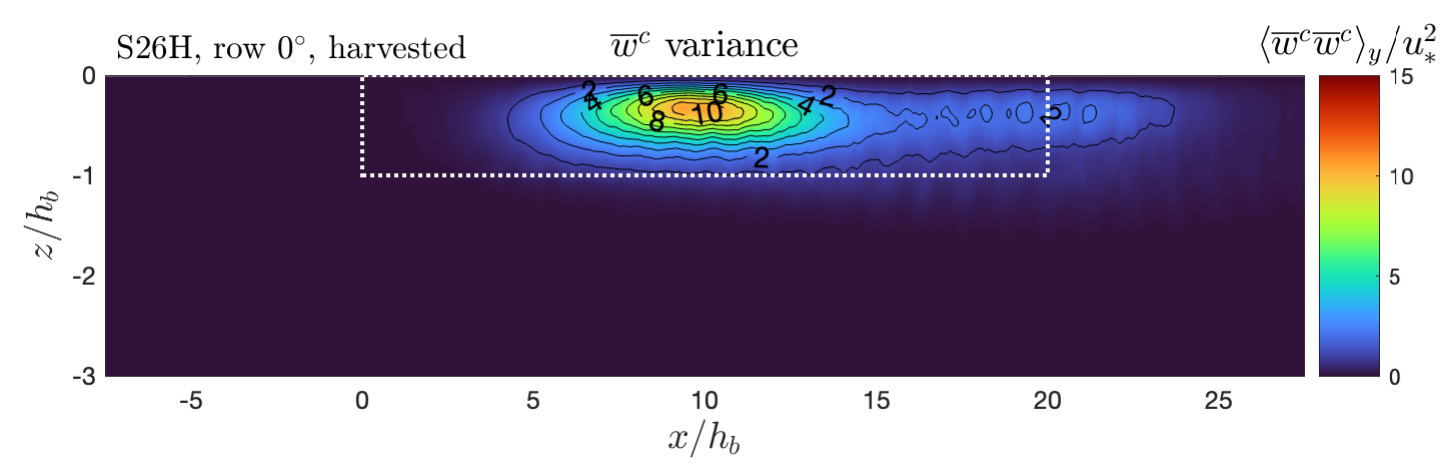}}
  \caption{Side view of the secondary flow component of vertical velocity 
  variance $\left\langle\overline{w}^c \overline{w}^c\right\rangle_{y}/u_*^2$ 
  for cases S26H. The results are temporally and laterally averaged. 
  The dotted rectangle shows the extent of the farm.}
\label{fig:sideview-wc2}
\end{figure}

In figure~\ref{fig:sideview-wc2}, we plot the temporally and laterally 
averaged vertical velocity variance associated with the steady secondary 
flow, denoted as $\left\langle\overline{w}^c\overline{w}^c\right\rangle_{y}$. 
Note that secondary flow includes both the lateral variations in 
streamwise velocity ($\overline{u}^c$) and the stationary lateral 
circulation ($\overline{v}^c$ and $\overline{w}^c$). 
Secondary flow exclusively occurs in the case with laterally 
spaced kelp rows (case S26H), where attached Langmuir circulation occurs. 
In farm block simulations where transient Langmuir circulation has been 
found, $\left\langle\overline{w}^c\overline{w}^c\right\rangle_{y}$ is 
generally negligible. The region of high 
$\left\langle\overline{w}^c\overline{w}^c\right\rangle_{y}$ roughly 
coincides with that of high $\left\langle\overline{w'w'}\right\rangle_{y}$ 
for case S26H, while the peak of $\left\langle\overline{w}^c\overline{w}^c
\right\rangle_{y}$ is slightly upstream of $\left\langle\overline{w'w'}
\right\rangle_{y}$ (figure~\ref{fig:along-wp2-wc2}). 

\begin{figure}
  \centerline{\includegraphics[width=0.7\textwidth]{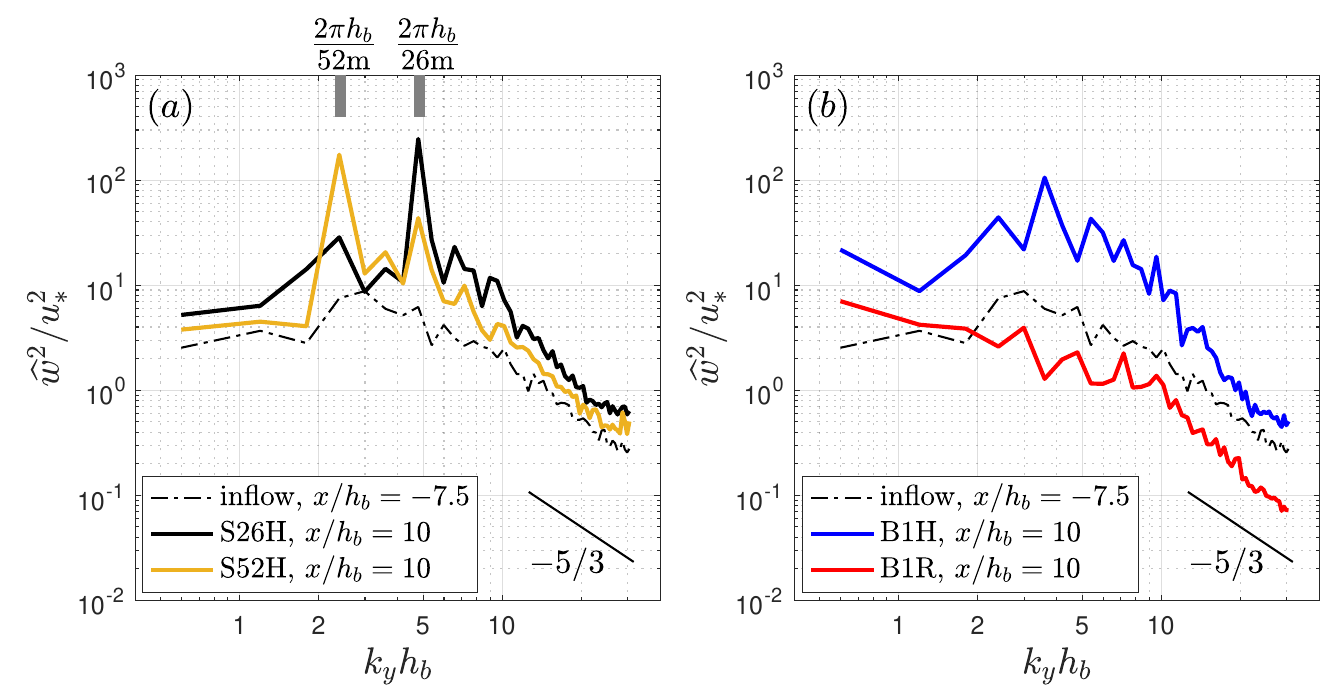}}
  \caption{Spectra of the vertical velocity 
  versus the lateral-direction wavenumber $k_y$. 
  $(a)$: Cases S26H (rows aligned with the $x$-direction, 26~m spacing, 
  solid black line) and S52H (52~m spacing, solid yellow line), 
  at $x/h_b=10$ and averaged between $z=0$ and $z=-h_b$. 
  The dash-dotted black line represents the inflow condition at 
  the upstream boundary ($x/h_b=-7.5$). 
  Two vertical gray lines indicate the corresponding spacing 
  between kelp rows, 26~m and 52~m, respectively. 
  $(b)$: Cases B1H (solid blue line) and B1R (solid red line).}
\label{fig:spectra}
\end{figure}

We calculated the lateral-direction wavenumber spectra of the 
vertical velocity to investigate the characteristic lateral 
spacing of Langmuir circulations. In case S26H (rows with a 
spacing of 26~m), the spectral peak in the farm aligns with the lateral row 
spacing $S_{MF}=26$~m (figure~\ref{fig:spectra}$a$). 
Similarly, for case S52H with rows spaced at 52~m, the  
peak wavelength corresponds to $S_{MF}=52$ m. Note that these spectra are 
calculated from the total vertical velocity $w$, and the energy 
peaks in the above two cases are predominantly contributed by 
the secondary flow component $\overline{w}^c$. 
The peaks would be considerably less distinct if calculating 
the spectra using the turbulent component $w'$ alone. 
In case B1H (farm block with the harvested profile), the 
spectral peak corresponds to a wavelength of around 30~m 
(figure~\ref{fig:spectra}$b$). This peak wavelength is slightly 
smaller than that of the inflow condition with standard 
Langmuir turbulence. In contrast to the spaced rows, the farm 
block exhibits a less distinct peak, due to the absence of 
secondary flow structures caused by row spacing. 
For Case B1R (farm block with the ripe profile), 
the spectrum exhibits no apparent peak because of the inhibition of 
Langmuir turbulence in the farm, with a weakened overall magnitude 
compared to that upstream of the farm. 

\subsection{Vertical velocity skewness}
\label{sec:results-skewness}
\begin{figure}
  \centerline{\includegraphics[width=0.8\textwidth]{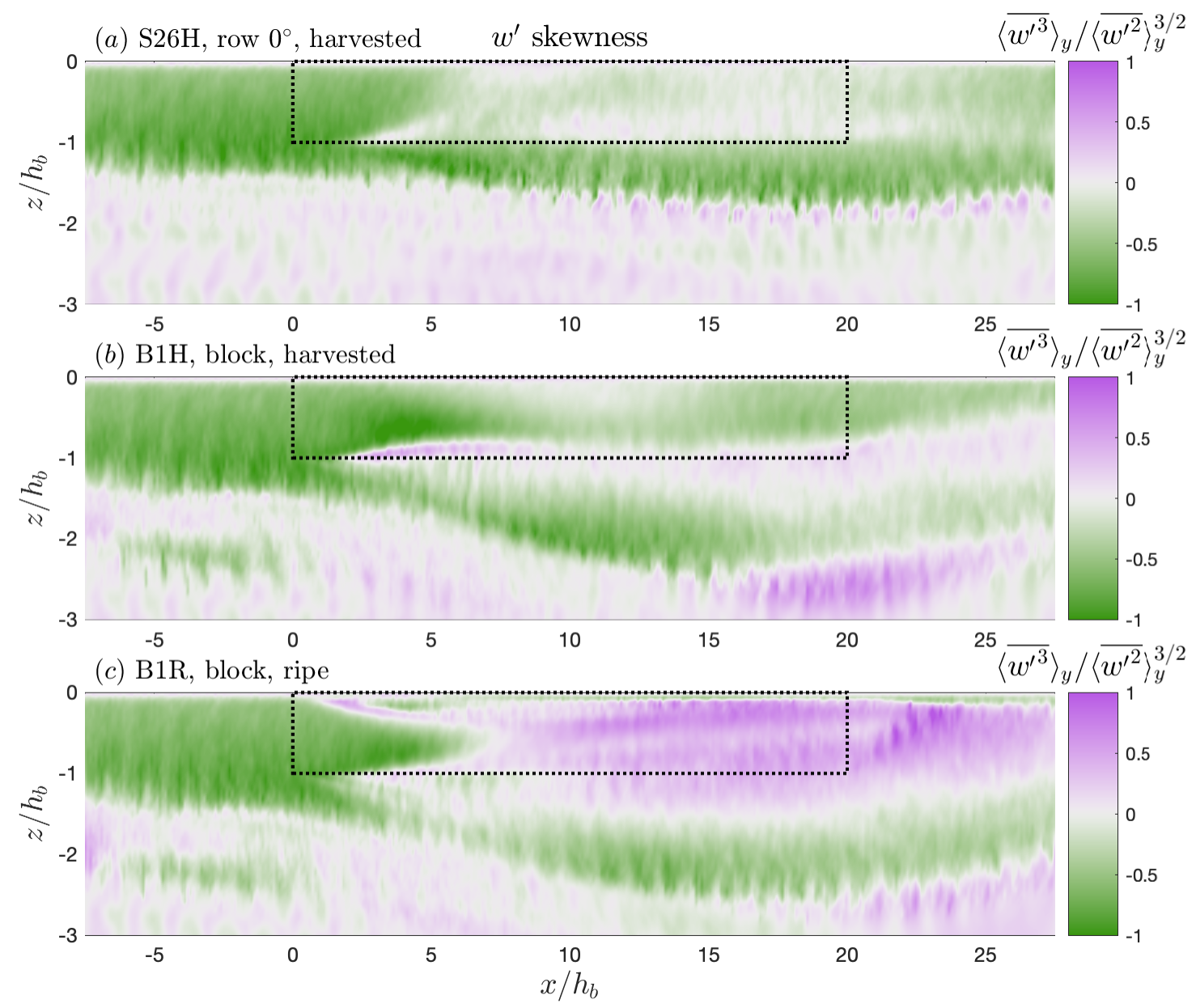}}
  \caption{Side views of the skewness of $w'$ (transient component 
  of vertical velocity) for cases S26H $(a)$, B1H $(b)$, and B1R $(c)$. 
  The results are temporally and laterally averaged. 
  Dotted rectangles show the extent of the farm.}
\label{fig:sideview-skewness}
\end{figure}

The skewness of the vertical turbulence component, defined as $\left\langle 
\overline{{w'}^3} \right\rangle_y/\left\langle\overline{{w'}^2}
\right\rangle_y^{3/2}$, is plotted in figure~\ref{fig:sideview-skewness} 
to illustrate the distinct attributes of various types of turbulence. 
The standard Langmuir turbulence in the upstream region is characterized 
by its negative skewness of vertical velocity \citep{mcwilliams1997}. 
The negative skewness indicates stronger downward motions confined within 
narrower regions in comparison to broader and weaker upward motions, 
consistent with the Langmuir circulation patterns shown in figure~\ref{fig:map-w}. 

In the shear layer generated at the canopy bottom, the vertical 
velocity skewness is generally negative beneath the canopy and 
positive near the canopy edge (figure~\ref{fig:sideview-skewness}). 
The positive skewness of vertical velocity near the canopy edge 
indicates the dominance of sweep events that bring high-momentum 
fluid into the canopy, consistent with those found in classical 
canopy flow without waves \citep{raupach1981,katul1997,poggi2004}. 
Note that the skewness sign in the suspended canopy is opposite to 
that of submerged benthic canopies due to the reversed geometry 
of the problem setups (i.e., here sweep events are characterized 
by intensified positive vertical velocities into the canopy). 
Conversely, the negative skewness of vertical velocity below the 
canopy is indicative of the prevalence of ejection events. 

The vertical velocity skewness of the farm-generated Langmuir 
turbulence appears less clear, displaying mostly negative 
values. However, the boundary between shear layer 
turbulence and farm-generated Langmuir turbulence remains 
discernible through their difference in the skewness sign 
inside the canopy. Specifically, 
in case B1H in figure~\ref{fig:sideview-skewness}$(b)$, the 
positive vertical velocity skewness associated with shear layer 
turbulence occupies less than one third of the canopy height. 
Furthermore, the region with the highest turbulence intensity 
is located above the region with positive skewness 
(figure~\ref{fig:sideview-wp2}$b$), underscoring that the 
majority of energy within the farm is attributed to Langmuir-type 
turbulence rather than the penetrated shear layer turbulence. 

For case B1R (farm block with the ripe profile), the concept of 
skewness within the farm lacks significance due to the absence of 
Langmuir-type turbulence. However, indications of shear-generated 
turbulence can be found within the farm, e.g., in the range of 
$x/h_b$ from 0 to 5, where the mean flow is adjusting to the canopy 
drag (figure~\ref{fig:sideview-skewness}$c$). This shear-generated 
turbulence within the farm exhibits similar characteristics to the 
shear layer turbulence below the canopy bottom edge, although with 
lower intensity. It results from the vertical variability in frond 
area density, and then rapidly dissipates downstream in the farm. 

\begin{figure}
  \centerline{\includegraphics[width=0.8\textwidth]{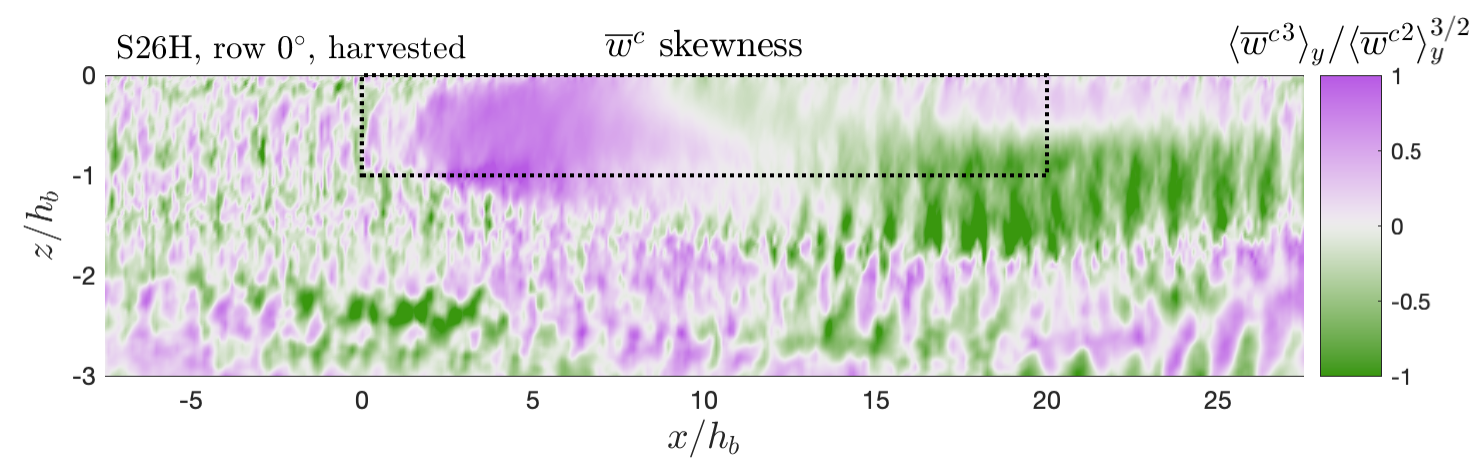}}
  \caption{Side view of the skewness of $\overline w^c$ (steady 
  secondary flow component of vertical velocity) for case S26H. 
  The results are temporally and laterally averaged. 
  The dotted rectangle shows the extent of the farm.}
\label{fig:sideview-skewness-circ}
\end{figure}

Similarly, we define a skewness for the steady secondary flow as 
$\left\langle {\overline w^c}^3\right\rangle_y/\left\langle 
{\overline w^c}^2\right\rangle_y^{3/2}$. The steady secondary 
velocity is generally upward in the kelp rows and downward in the 
gaps between kelp rows (figure~\ref{fig:map-w}). In cases where 
the kelp rows are narrower than the gaps, e.g., case S26H in 
figure~\ref{fig:sideview-skewness-circ}, the upward secondary flow 
is typically stronger in magnitude than the downward flow as a 
result of mass balance. Therefore, a positive skewness of the 
steady secondary vertical velocity is evident in the upstream 
part of the farm (figure~\ref{fig:sideview-skewness-circ}). 
Toward the downstream region of the farm, the vertical secondary 
velocity skewness becomes less distinct. This can be attributed to 
the lateral expansion of Langmuir wakes linked to kelp rows, so that 
regions of upwelling may extend beyond the constraints of kelp rows 
(figure~\ref{fig:map-w}$a$, also see Appendix~\ref{app-single-row} for 
a clearer presentation of the Langmuir wake pattern in case S208H, 
the single-row simulation). 
As one may anticipate, the skewness of secondary flow  
varies with changes in farm width and lateral spacing 
across different farm configurations. As an example, the skewness 
$\left\langle {\overline w^c}^3\right\rangle_y/\left\langle 
{\overline w^c}^2\right\rangle_y^{3/2}$ in case S208H can reach 
up to a value of 3, where the upwelling of secondary flow is confined within 
a single kelp row before wake expansion occurs. 

The skewness of turbulence and secondary flow investigated above 
also influences diffusive and dispersive fluxes in the farm, 
potentially leading to asymmetric vertical nutrient transport 
\citep{wyngaard1984,wyngaard1991,chor2020,chor2021}. 
Note that spurious oscillations displaying alternating 
positive and negative skewness can be found in the secondary 
flow and turbulence components, e.g., around $z/h_b=-2$ and 
$x/h_b=15$ in figure~\ref{fig:sideview-skewness-circ} and around 
$z/h_b=-2$ and $x/h_b=20$ in figure~\ref{fig:sideview-skewness}$(a)$. 
These oscillations occur in regions where $\overline w^c$ 
or $w'$ is minimal and thus do not have practical significance.

\section{Kinetic energy balance}
\label{sec:tke-budget}
In this section we investigate the kinetic energy equations to understand the 
sources of Langmuir circulation and turbulence. Following the convention 
in \eqref{eq:vel-decomp} and \eqref{eq:flux-decomp}, we decompose the 
total kinetic energy $K_T$ into the mean flow energy $K_M$, secondary 
flow energy $K_{SE}$, and turbulence kinetic energy $K_{TE}$, i.e., 
\begin{linenomath*}
\begin{equation}
    \label{eq:tke-component}
    K_T = \frac{1}{2}\left\langle \overline{u_i u_i}\right\rangle_y 
    = \underbrace{\frac{1}{2}\left\langle \overline{u_i}\right\rangle_y 
    \left\langle\overline{u_i}\right\rangle_y 
    \vphantom{\frac{1}{2}\left\langle \overline{u'_i u'_i}\right\rangle_y} }_{K_M}
    + \underbrace{\frac{1}{2}\left\langle \overline{u_i}^c 
    \overline{u_i}^c \right\rangle_y 
    \vphantom{\frac{1}{2}\left\langle \overline{u'_i u'_i}\right\rangle_y} }_{K_{SE}}
    + \underbrace{\frac{1}{2}\left\langle \overline{u'_i u'_i} 
    \right\rangle_y}_{K_{TE}}.
\end{equation}
\end{linenomath*}

We can derive the transport equations for $K_M$, $K_{SE}$, and 
$K_{TE}$ from the governing equations \eqref{eq:mass} and \eqref{eq:mom} \citep{yan2021}:
\begin{linenomath*}
\begin{subeqnarray}
    \label{eq:tke-budget}
    \slabel{eq:tke-budget-KM}
    \Ldiff{K_M}{t} &=& - C_{M-SE} - C_{M-TE} + S_M 
                    + D_M + T_M + R_M, \\
    \slabel{eq:tke-budget-KSE}
    \Ldiff{K_{SE}}{t} &=& C_{M-SE} - C_{SE-TE} + S_{SE} 
                    + D_{SE} + T_{SE}, \\
    \slabel{eq:tke-budget-KTE}
    \Ldiff{K_{TE}}{t} &=& C_{M-TE} + C_{SE-TE} + S_{TE} + 
                    \epsilon_{TE} + D_{TE} + T_{TE}.
\end{subeqnarray}
\end{linenomath*}
Here the material derivative is $\text{D}/\text{D}t = \partial/\partial t 
+ \left\langle\overline{u_j}\right\rangle_y\partial/\partial x_j + 
u_s\partial/\partial x$. Note that we have neglected the buoyancy 
production terms, as the farm is located within the OML. 
Terms $C_{M-SE}$, $C_{M-TE}$, and $C_{SE-TE}$ 
represent the energy conversion between the mean flow $K_M$, secondary 
flow $K_{SE}$, and turbulence $K_{TE}$ 
(also referred to as shear production), written as  
\begin{linenomath*}
\begin{subeqnarray}
    \label{eq:tke-conversion}
    C_{M-SE} &=& - \left\langle \overline{u_i}^c \overline{u_j}^c\right\rangle_y 
                    \pdiff{\left\langle\overline{u_i}\right\rangle_y}{x_j}, \\
    C_{M-TE} &=& - \left\langle \overline{u_i' u_j'}\right\rangle_y 
                    \pdiff{\left\langle\overline{u_i}\right\rangle_y}{x_j}, \\
    C_{SE-TE} &=& - \left\langle \overline{u_i' u_j'} 
                    \pdiff{\overline{u_i}^c}{x_j}\right\rangle_y.
\end{subeqnarray}
\end{linenomath*}
Terms $S_{M}$, $S_{SE}$, and $S_{TE}$ stand for Stokes production 
that transfers energy between waves and the three flow components, and 
are given by 
\begin{linenomath*}
\begin{subequations}
    \begin{gather}\label{eq:tke-stokes}
    S_{M} = - \left\langle \overline{u}\right\rangle_y 
                \left\langle \overline{w}\right\rangle_y \pdiff{u_s}{z}, \quad 
    S_{SE} = - \left\langle \overline{u}^c \overline{w}^c 
                \right\rangle_y \pdiff{u_s}{z}, \quad 
    S_{TE} = - \left\langle \overline{u'w'}
                \right\rangle_y \pdiff{u_s}{z}. \tag{\theequation a-c}
    \end{gather}
\end{subequations}
\end{linenomath*}
The SGS dissipation term is 
\begin{linenomath*}
\begin{equation}
    \label{eq:tke-dissip}
    \epsilon_{TE} = -\left\langle \overline{\tau'_{ij} 
                \pdiff{u'_i}{x_j}}\right\rangle_y.
\end{equation}
\end{linenomath*}
The majority of energy dissipation is expected to occur in the small-scale 
transient eddies for turbulent flows with a large Reynolds number 
\citep[e.g.,][]{pope2000}. Therefore, the energy loss of the 
larger-scale mean flow and secondary flow due to direct SGS dissipation is 
considered negligible, and we only focus on the dissipation of turbulence 
kinetic energy. 
The canopy drag dissipation terms are 
\begin{linenomath*}
\begin{subequations}
    \begin{gather}\label{eq:tke-drag}
    D_{M} = -\left\langle\overline{u_i}\right\rangle_y 
                \left\langle\overline{F_{D,i}}\right\rangle_y, \quad 
    D_{SE} = -\left\langle \overline{u_i}^c 
                \overline{F_{D,i}}^c\right\rangle_y, \quad 
    D_{TE} = -\left\langle \overline{u'_{i} F'_{D,i}}\right\rangle_y, 
    \tag{\theequation a-c}
    \end{gather}
\end{subequations}
\end{linenomath*}
which represent the energy loss or gain associated with canopy drag. 
In addition, several flux terms representing kinetic energy transport 
associated with the resolved stress, SGS stress, and pressure 
are collected into the transport terms $T_M$, $T_{SE}$, and $T_{TE}$. 
The last term $R_M= f \left\langle\overline{v}\right\rangle_y (u_g-u_s)$ 
in the mean flow energy budget represents the energy transfer between 
the mean flow, surface waves and the background geostrophic current, 
resembling the concept of Stokes-Coriolis work \citep{suzuki2016,yan2021}.

\begin{figure}
  \centerline{\includegraphics[width=0.75\textwidth]{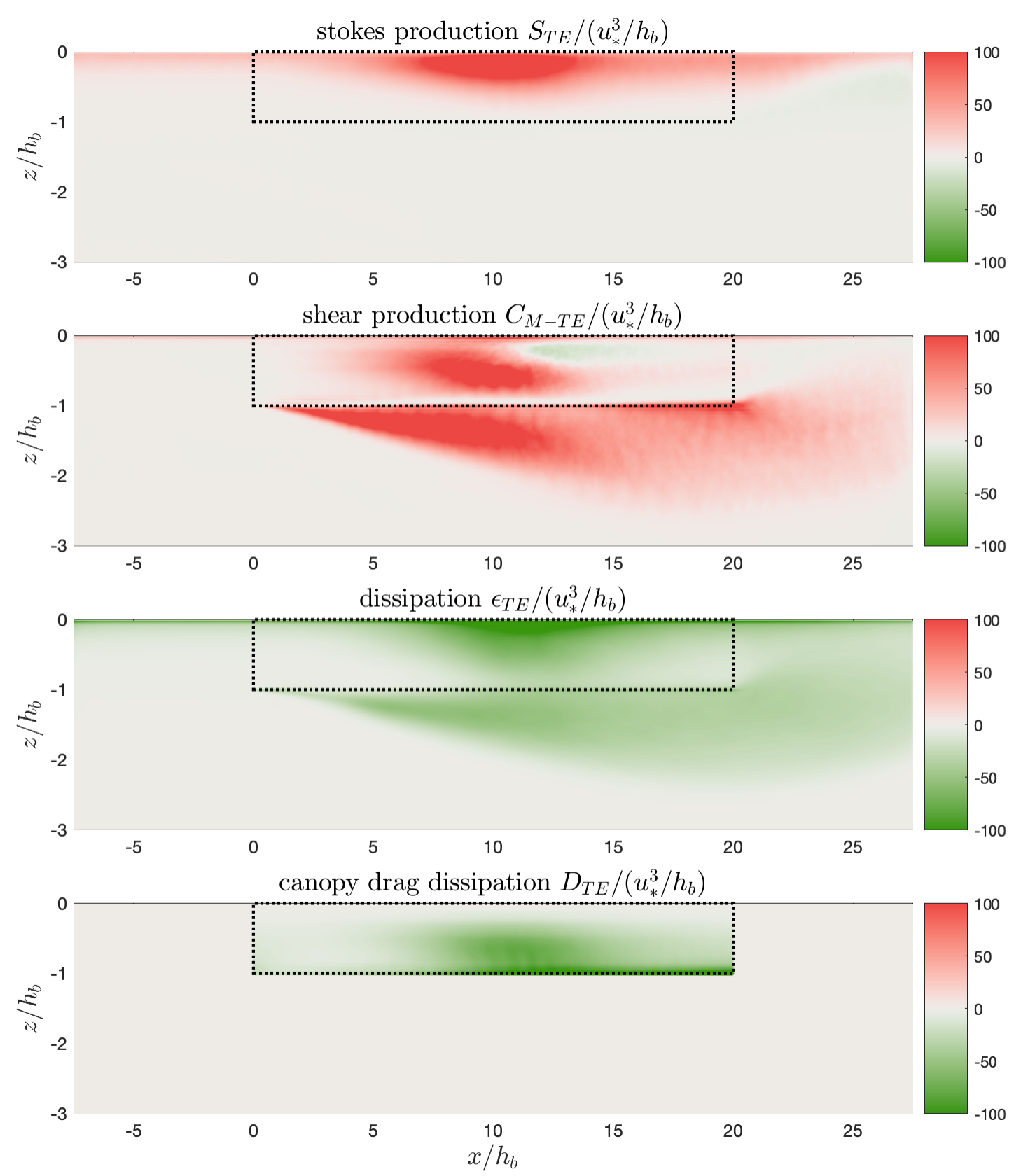}}
  \caption{Side views of terms in the $K_{TE}$ budget in 
  \eqref{eq:tke-budget-KTE} for case B1H (farm block, harvested profile). 
  The results are temporally and laterally averaged. 
  Dotted rectangles show the extent of the farm.}
\label{fig:sideview-tke-1}
\end{figure}

We calculate the turbulent kinetic energy budget for case B1H (farm block 
with the harvested profile), and the major source and sink terms are 
shown in figure~\ref{fig:sideview-tke-1}. The Stokes production is the 
dominant source of turbulence upstream of the farm, where standard Langmuir 
turbulence is found. This agrees with the widely recognized mechanism 
wherein the generation of Langmuir circulation results from vorticity 
tilting by the Stokes drift \citep{craik1977,leibovich1977,leibovich1983}. 
Within the farm, the Stokes production becomes significantly stronger, 
which, together with the shear production, contributes to the enhanced 
Langmuir-type turbulence  (figures~\ref{fig:sideview-tke-1} and 
\ref{fig:sideview-wp2}). Below the farm, shear production provides 
the dominant source of shear layer turbulence, and Stokes production 
is weak because the Stokes drift diminishes at greater depth. Both  
dissipation and canopy drag dissipation terms tend to destroy the turbulence 
generated within the farm, while in the shear layer below the canopy, the 
dissipation term is the only sink of turbulence kinetic energy. 
Note that the secondary flow energy is negligible in the farm block, 
and there is no energy conversion associated with secondary flow. 

\begin{figure}
  \centerline{\includegraphics[width=0.75\textwidth]{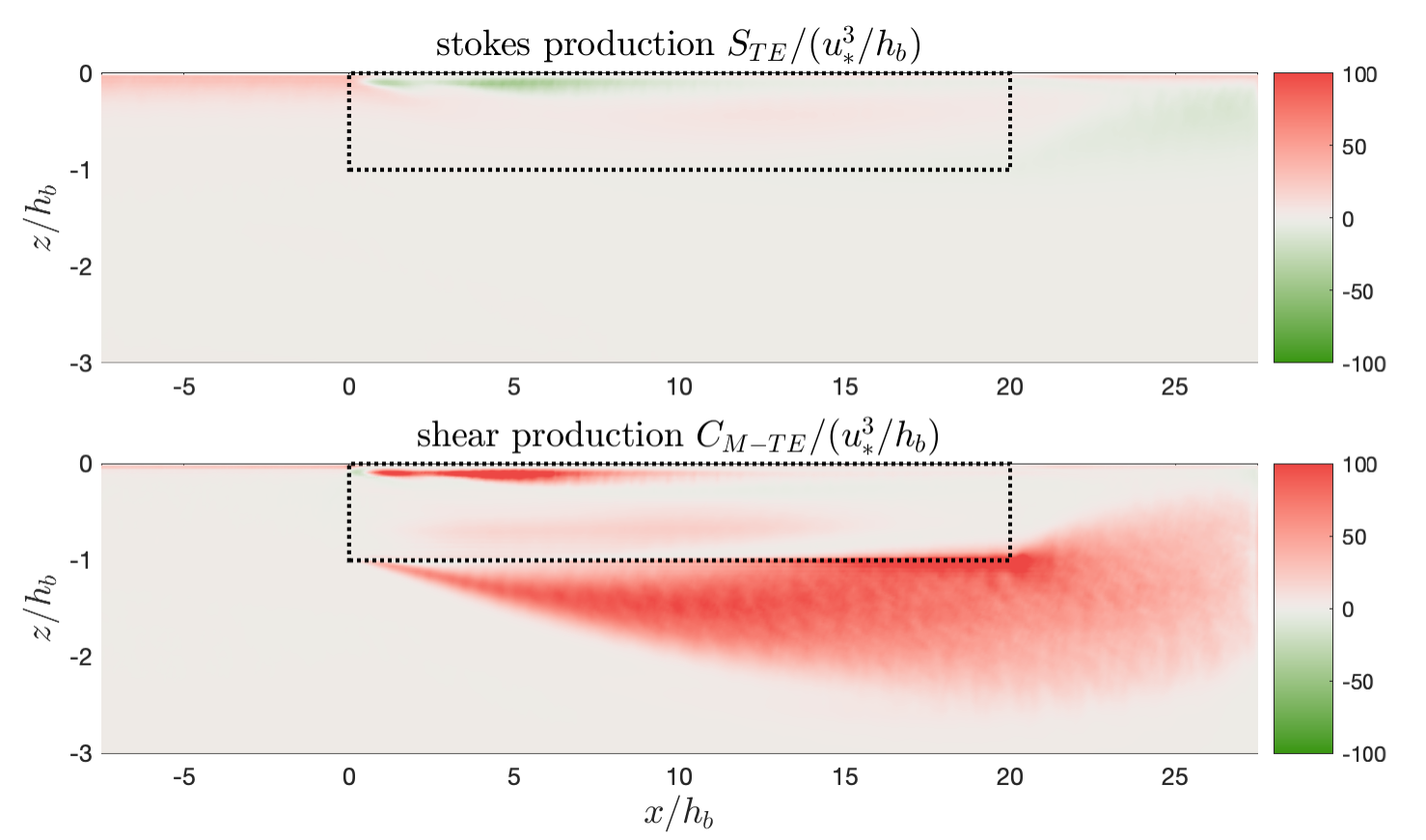}}
  \caption{Side views of terms in the $K_{TE}$ budget in 
  \eqref{eq:tke-budget-KTE} for case B1R (farm block, ripe profile). 
  The results are temporally and laterally averaged. 
  Dotted rectangles show the extent of the farm. 
  Note that differences between case B1R and case B1H 
  (figure~\ref{fig:sideview-tke-1}) primarily lie in the 
  energy production mechanisms. 
  The dissipation and canopy drag dissipation terms are 
  not shown here, because the energy loss mechanisms in case 
  B1R are similar to case B1H.}
\label{fig:sideview-tke-2}
\end{figure}

In case B1R (farm block with the ripe profile), the shear production 
term is the major source of turbulence kinetic energy in the shear 
layer below the canopy (figure~\ref{fig:sideview-tke-2}), consistent 
with case B1H discussed above (farm block with the harvested profile). 
However, when focusing within the farm, both the Stokes production 
and shear production diminish due to the presence of a dense layer 
near the surface (characteristic of the ripe profile). 
This explains the absence of Langmuir-type turbulence in 
figure~\ref{fig:sideview-wp2}$(c)$. 
In particular, negative values of Stokes production are even found 
near the sea surface, indicating that the shear of the Stokes drift 
velocity contributes to disrupting vortex formation rather than 
stretching vortex tubes (detailed mechanisms are explained in 
§~\ref{sec:vort-dynamics}). Negative vertical shear of the Eulerian velocity 
arises due to the presence of a dense kelp layer near the surface, 
which dominates over the positive vertical shear of the Stokes drift 
(figure~\ref{fig:mean-vel-prof}). This results in a negative sign for 
the Reynolds stress $\overline{u'w'}$, and consequently, the 
Stokes production term acts as a sink for turbulence kinetic energy. 
The dissipation and kelp drag dissipation terms are not shown for case B1R. 
The distribution of dissipation generally correlates with that of 
turbulence intensity, and strong kelp drag dissipation occurs where 
areas with high kelp frond density intersect with intense turbulence. 

It is worthwhile noting that the originally farm-generated vertical 
shear has a comparable magnitude in cases B1R and B1H, e.g., near 
the farm leading edge in the range of $x/h_b$ between 0 and 5 
(figure~\ref{fig:mean-vel-prof}). In fact, case B1R has even more 
pronounced sheared flow near the surface due to the existence of 
the dense layer. Nonetheless, the rapid downstream increase in 
shear production in case B1H compared to B1R, e.g., for $x/h_b$ 
from 5 to 15, indicates that its strength does not solely depend on 
the vertical shear induced by variations in frond area density. 
Instead, the increased Stokes production in case 
B1H plays a critical role in initiating the growth of turbulence 
and thus contributing to the enhancement of shear production. 
Further investigation of the distinct Stokes mechanisms in the two cases 
will be presented in §~\ref{sec:vort-dynamics}.

\begin{figure}
  \centerline{\includegraphics[width=0.75\textwidth]{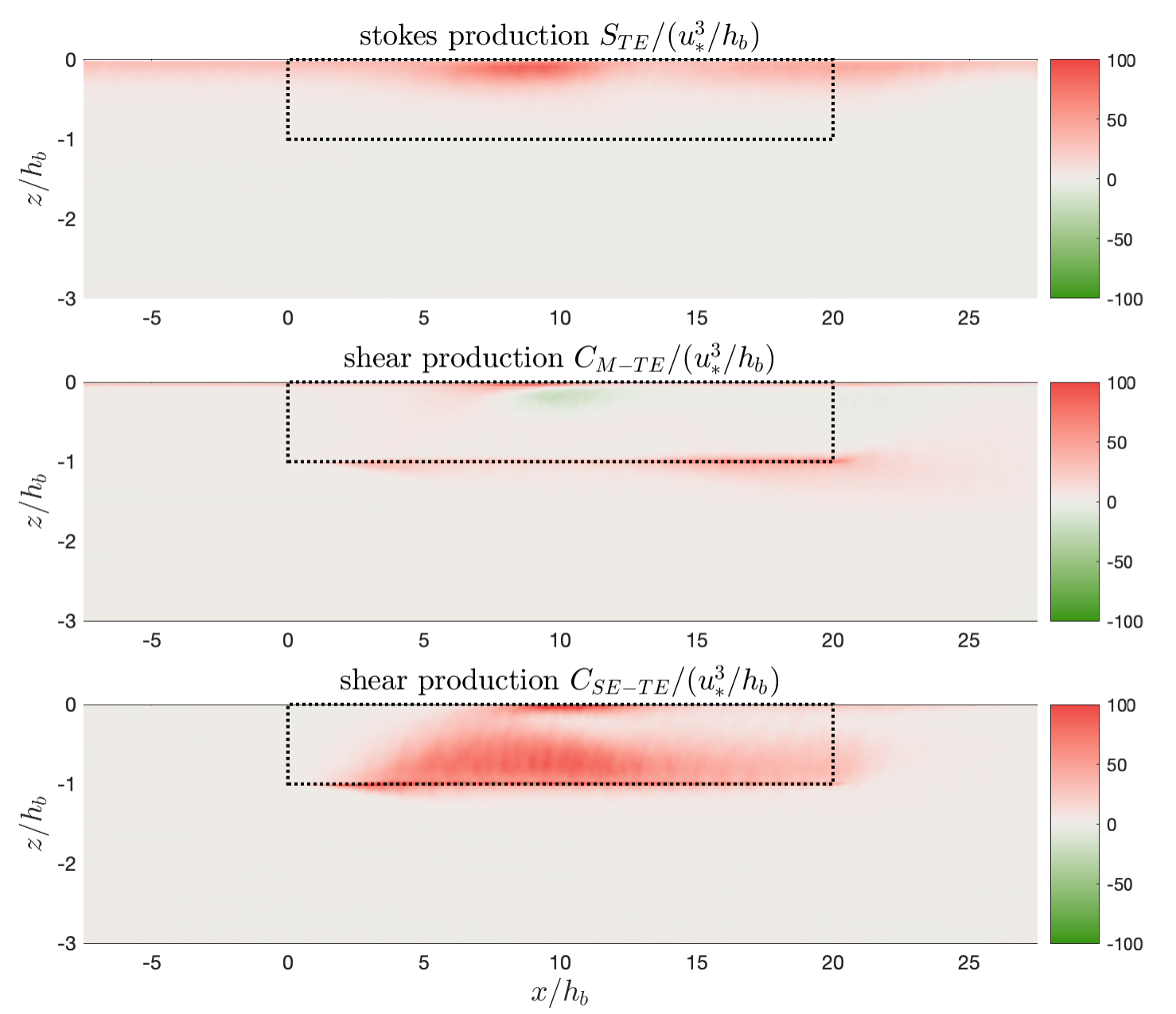}}
  \caption{Side views of terms in the $K_{TE}$ budget in \eqref{eq:tke-budget-KTE} 
  for case S26H (rows aligned with the current, harvested profile). 
  The results are temporally and laterally averaged. 
  Dotted rectangles show the extent of the farm.}
\label{fig:sideview-tke-3}
\end{figure}

The turbulence kinetic energy budget is also evaluated for the 
case with laterally spaced rows (S26H, harvested profile). The 
distributions of Stokes production (figure~\ref{fig:sideview-tke-3}), 
dissipation, and canopy drag dissipation (not shown) are generally 
similar to those in the farm block case (B1H, harvested profile, 
figure~\ref{fig:sideview-tke-1}), although with smaller magnitudes 
due to the overall lower effective density of the spaced rows. 
However, there is a distinction in shear production in the spaced 
rows as compared to the farm block, owing to the existence of 
secondary flow. Unlike the farm block case, where shear production 
directly converts mean flow energy to turbulence kinetic energy, 
a significant portion of turbulence kinetic energy in spaced rows 
is generated through the conversion of secondary flow energy. 

\begin{figure}
  \centerline{\includegraphics[width=0.75\textwidth]{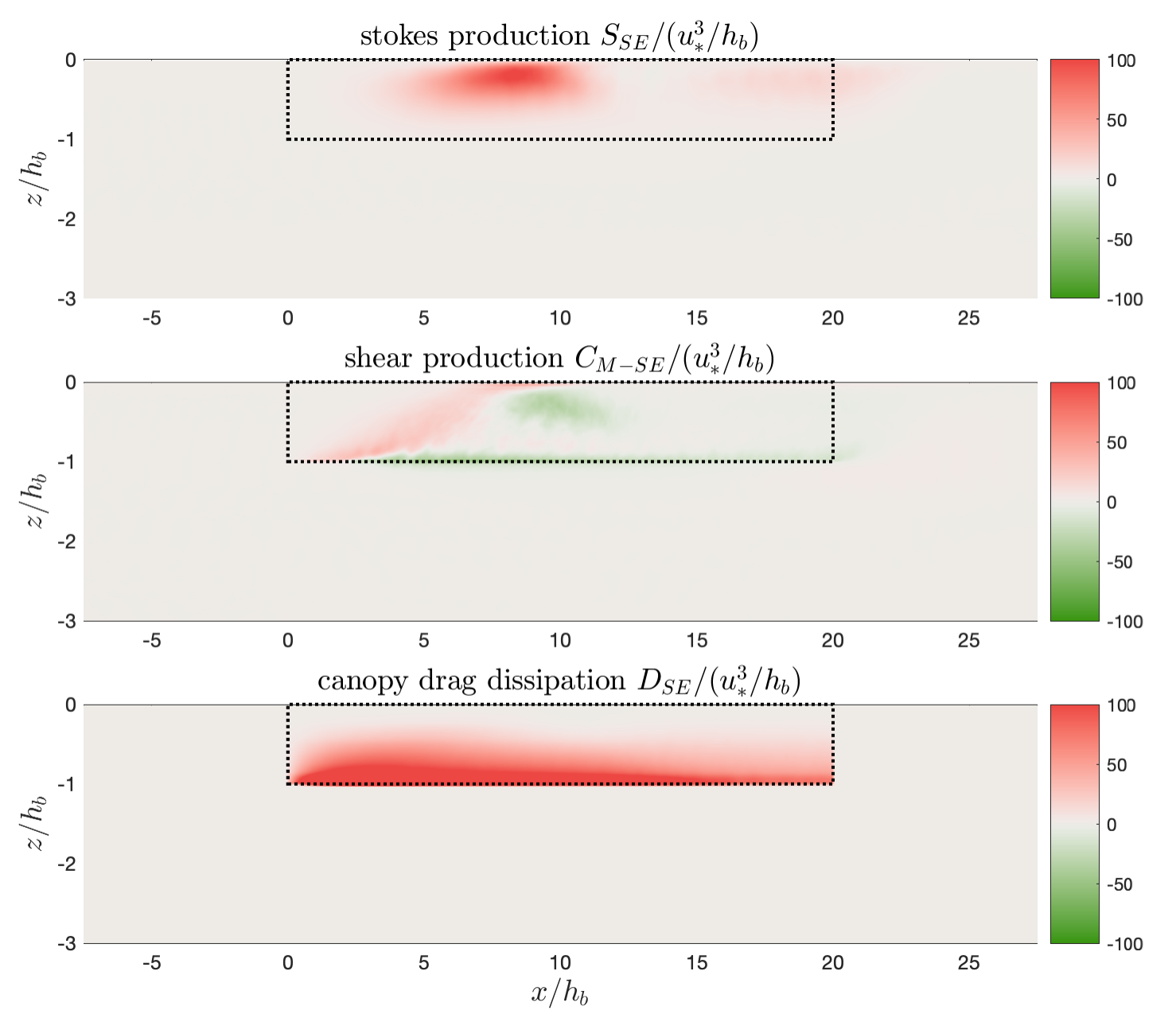}}
  \caption{Side views of terms in the $K_{SE}$ budget in \eqref{eq:tke-budget-KSE} 
  for case S26H (rows aligned with the current, harvested profile). 
  The results are temporally and laterally averaged. 
  Dotted rectangles show the extent of the farm.}
\label{fig:sideview-tke-4}
\end{figure}

We subsequently calculate the secondary flow energy budget 
(figure~\ref{fig:sideview-tke-4}). Note that the secondary flow 
energy encompasses both cross-stream variations in streamwise 
velocity ($\overline{u}^c$) and the steady lateral circulation 
($\overline{v}^c$ and $\overline{w}^c$). 
Canopy dissipation $D_{SE}$ is a significant source term in the 
secondary flow energy budget. Although the canopy drag leads to 
energy loss for both the mean flow and turbulence (hence named as 
a dissipation term), it can lead to a gain in secondary flow energy. 
This energy gain through $D_{SE}$ is evident as cross-stream 
variations in the drag force, enhancing the cross-stream 
variability of streamwise flow. In addition, the Stokes production term is 
another major source of secondary flow energy, contributing to 
the generation of attached Langmuir circulation.

\section{Vorticity dynamics of Langmuir circulations}
\label{sec:vort-dynamics}
The vorticity dynamics is investigated to understand the generation 
mechanisms of Langmuir circulation associated with various farm 
configurations. By manipulating the governing equations \eqref{eq:mass} and 
\eqref{eq:mom}, we derive the transport equation for vorticity 
$\boldsymbol{\zeta}$ \citep[e.g.,][]{fujiwara2018}:
\begin{linenomath*}
\begin{eqnarray}
\label{eq:vort-vec}
\pdiff{\boldsymbol{\zeta}}{t} +(\boldsymbol{u}\cdot\nabla)\boldsymbol{\zeta} 
    + &&(\boldsymbol{u}_s\cdot\nabla)\boldsymbol{\zeta} = 
     (\boldsymbol{\zeta}\cdot\nabla)\boldsymbol{u} 
     + (\boldsymbol{\zeta}\cdot\nabla)\boldsymbol{u}_s 
     - \nabla\times\boldsymbol{F}_D   \nonumber \\
     && + \nabla\times(\nabla\cdot\boldsymbol{\tau}^d) 
     - \nabla\times\frac{\rho}{\rho_0} g\boldsymbol{e}_z 
     + f\boldsymbol{e}_z\cdot\nabla(\boldsymbol{u}+\boldsymbol{u}_s).
\end{eqnarray}
\end{linenomath*}
The time derivative and advection terms are on the left side. 
The first two terms on the right side represent vorticity tilting 
and Stokes drift-vorticity tilting, respectively. The third term 
corresponds to the vorticity generation associated with canopy drag. 
The fourth to sixth terms represent the contributions of SGS stress, 
baroclinicity, and rotation. For the sake of clarity, we express 
the equations for each vorticity component separately 
\begin{linenomath*}
\begin{subeqnarray}
\label{eq:vort-xyz}
\slabel{eq:vort-x}
\Ldiff{\zeta_x}{t} &=& \zeta_x\pdiff{u}{x} + \zeta_y\pdiff{u}{y} 
+ \zeta_z\pdiff{u}{z} + \zeta_z\pdiff{u_s}{z} + R_{\zeta,x}, \\
\slabel{eq:vort-y}
\Ldiff{\zeta_y}{t} &=& \zeta_x\pdiff{v}{x} + \zeta_y\pdiff{v}{y} 
+ \zeta_z\pdiff{v}{z} - \pdiff{F_{D,x}}{z}+ R_{\zeta,y}, \\
\slabel{eq:vort-z}
\Ldiff{\zeta_z}{t} &=& \zeta_x\pdiff{w}{x} + \zeta_y\pdiff{w}{y} 
+ \zeta_z\pdiff{w}{z} + \pdiff{F_{D,x}}{y}+ R_{\zeta,z}.
\end{subeqnarray}
\end{linenomath*}
The material derivative is defined as $\text{D}/\text{D}t = \partial/\partial t 
+ \left\langle\overline{u_j}\right\rangle_y\partial/\partial x_j + 
u_s\partial/\partial x$, the same as \eqref{eq:tke-budget}. 
Note that we have assumed $F_{D,x}\gg F_{D,y},F_{D,z}$, given that 
the streamwise velocity is much greater than lateral circulation, 
and we thus only focus on the influence of $F_{D,x}$ on vorticity. 
The term related to SGS stress diffuses vorticity rather than 
generating it; the baroclinic effect can be neglected because the 
farm is within the mixed layer; the effect of Earth rotation is minimal. 
We thus group these factors into the residual terms, denoted as 
$R_{\zeta,x}$, $R_{\zeta,y}$ and $R_{\zeta,z}$. 

The vorticity is calculated for case S26H (spaced kelp rows aligned with 
the current, harvested profile). Here we introduce the phase average 
in the cross-stream direction, denoted as $\langle\cdot\rangle_p$, to 
average over equivalent positions of the periodically repeated rows. 
For any field $\phi$, the cross-stream phase average is defined as 
\begin{linenomath*}
\begin{equation}
    \left\langle\phi\right\rangle_p(x,y,z) = \frac{1}{N}\sum_{n=0}^{N-1} 
    \phi\left(x,y+nS_{MF}+\frac{1}{2}W_{MF},z\right),
\end{equation}
\end{linenomath*}
where $N$ is the total number of rows. 

\begin{figure}
  \centerline{\includegraphics[width=1.1\textwidth]{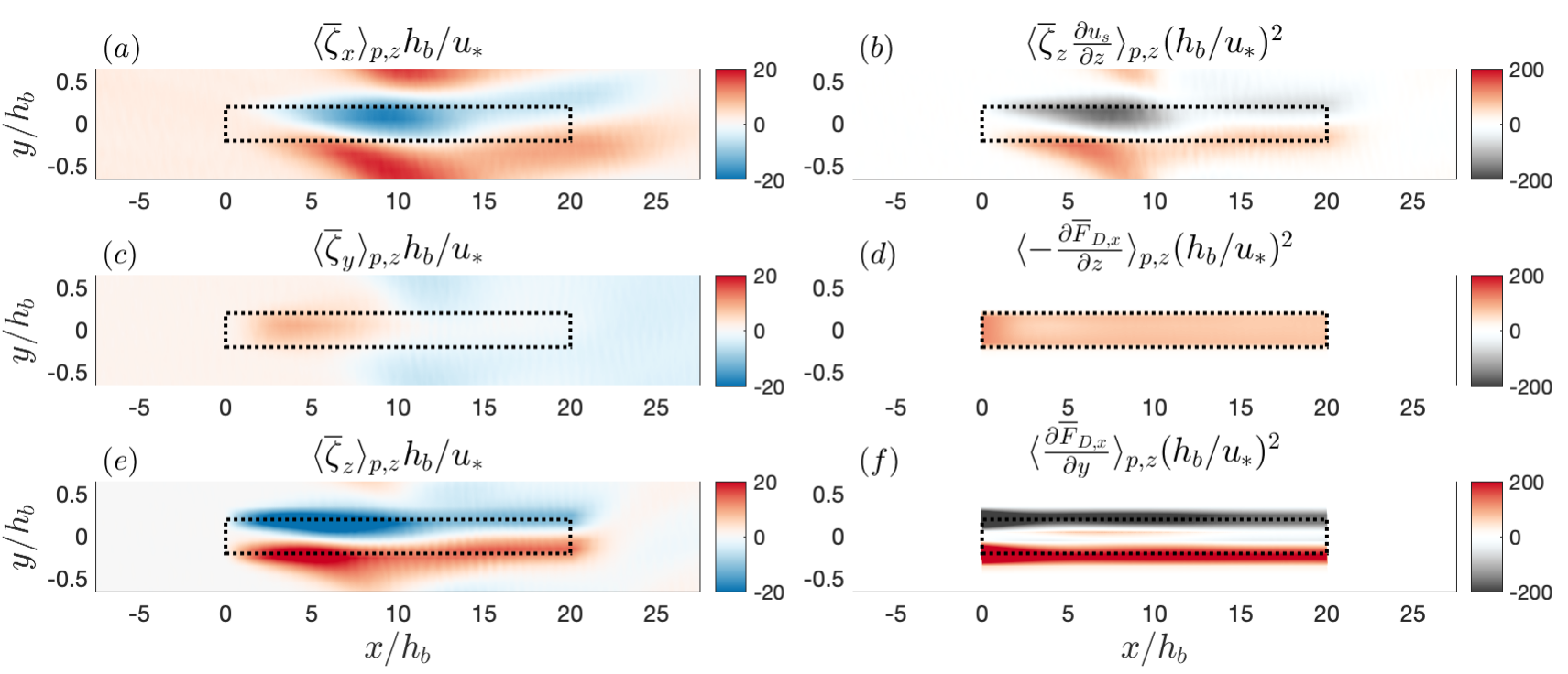}}
  \caption{Top views of vorticity components $\zeta_x$ ($a$), $\zeta_y$ ($c$), 
  $\zeta_z$  ($e$), and their primary forcing terms ($b$, $d$, and $f$) in \eqref{eq:vort-xyz} 
  for case S26H (rows aligned with the current, harvested profile). 
  The results are time-averaged, depth-averaged ($z=0$ to $-h_b$), 
  and cross-phase-averaged.  
  Dotted rectangles show the extent of the kelp row.}
\label{fig:map-vort-sp26}
\end{figure}

\begin{figure}
  \centerline{\includegraphics[width=0.8\textwidth]{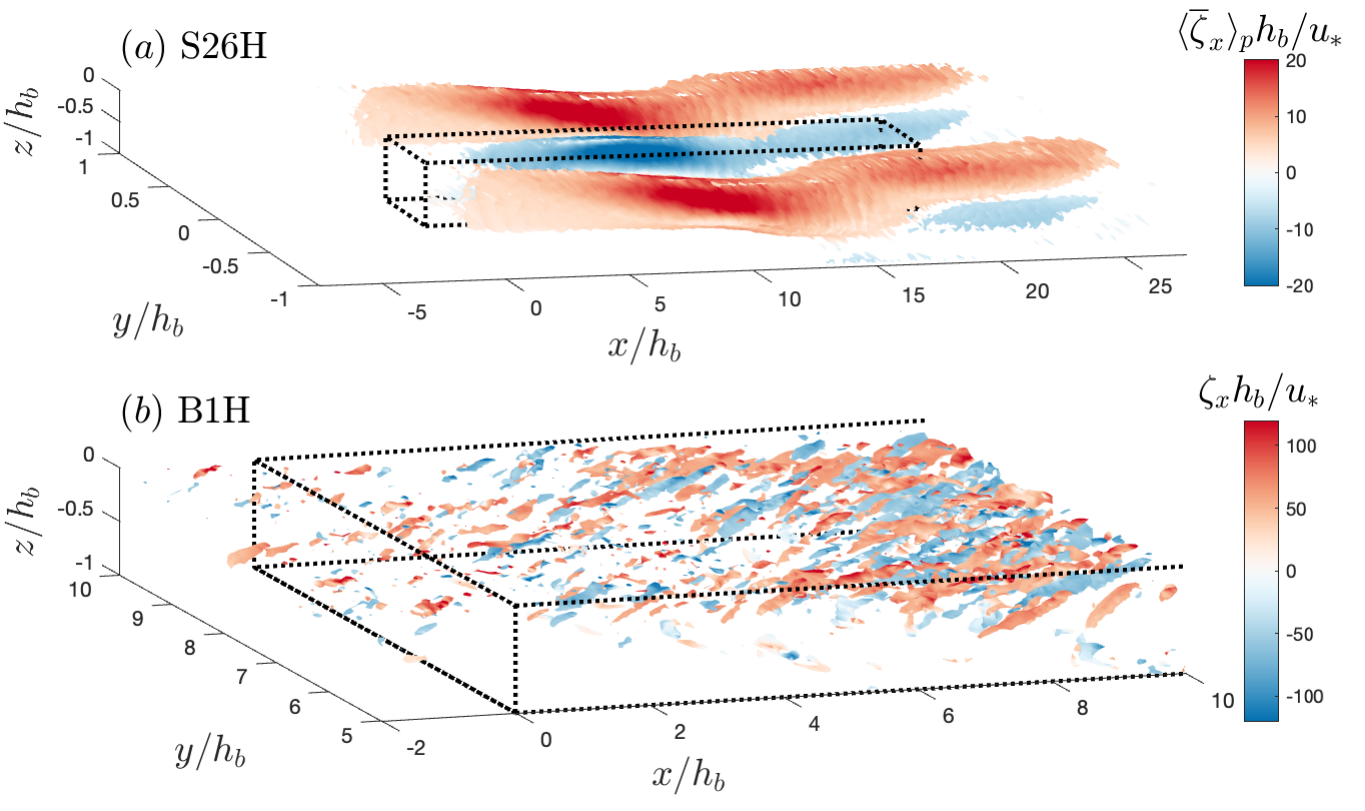}}
  \caption{Three-dimensional representation of vortex structures, 
  as revealed by isosurfaces of $\lambda_2$. 
  $(a)$: Cross-phase-averaged and time-averaged streamwise vorticity 
  $\left\langle\overline\zeta_x\right\rangle_p$ for a kelp row in case S26H 
  (rows aligned with the current, harvested profile). 
  $(b)$: A zoom-in view of the instantaneous streamwise vorticity $\zeta_x$ 
  for case B1H (farm block, harvested profile). Note the differences in 
  colormap range and axis range between the two plots.}
\label{fig:3d-vort}
\end{figure}

The dominant phase-averaged terms in the vorticity equation are shown 
in figure~\ref{fig:map-vort-sp26}. The persistent downstream vorticity 
$\zeta_x$ corresponds to the attached Langmuir circulation generated 
by kelp rows (recall figure~\ref{fig:map-w}$a$). The generation of 
attached Langmuir circulation is attributed to the tilting of vertical 
vorticity $\zeta_z$ by the Stokes drift into the downstream direction, 
i.e., the term $\zeta_z (\partial u_s/\partial z)$ in \eqref{eq:vort-x}. 
This tilting process aligns with the classical mechanism that 
gives rise to standard Langmuir circulation \citep[e.g.,][]{leibovich1983}. 
In the farm the vertical vorticity $\zeta_z$ is predominantly 
generated due to the drag discontinuity at the lateral edges of kelp 
rows, i.e., $\partial F_{D,x}/\partial y$ in \eqref{eq:vort-z}. 
The enhanced lateral shear in streamwise velocity ($\zeta_z$) 
due to spaced kelp rows can thus lead to more intensified Langmuir 
circulation ($\zeta_x$) within the farm (figures~\ref{fig:map-vort-sp26} 
and \ref{fig:3d-vort}$a$), compared to the standard 
Langmuir circulation upstream of the farm \citep{yan2021}. 
Note that the vorticity field in figure~\ref{fig:map-vort-sp26} 
was averaged over time to show the persistent forcing 
and stationary Langmuir circulation associated with 
spaced kelp rows. 

\begin{figure}
  \centerline{\includegraphics[width=1.1\textwidth]{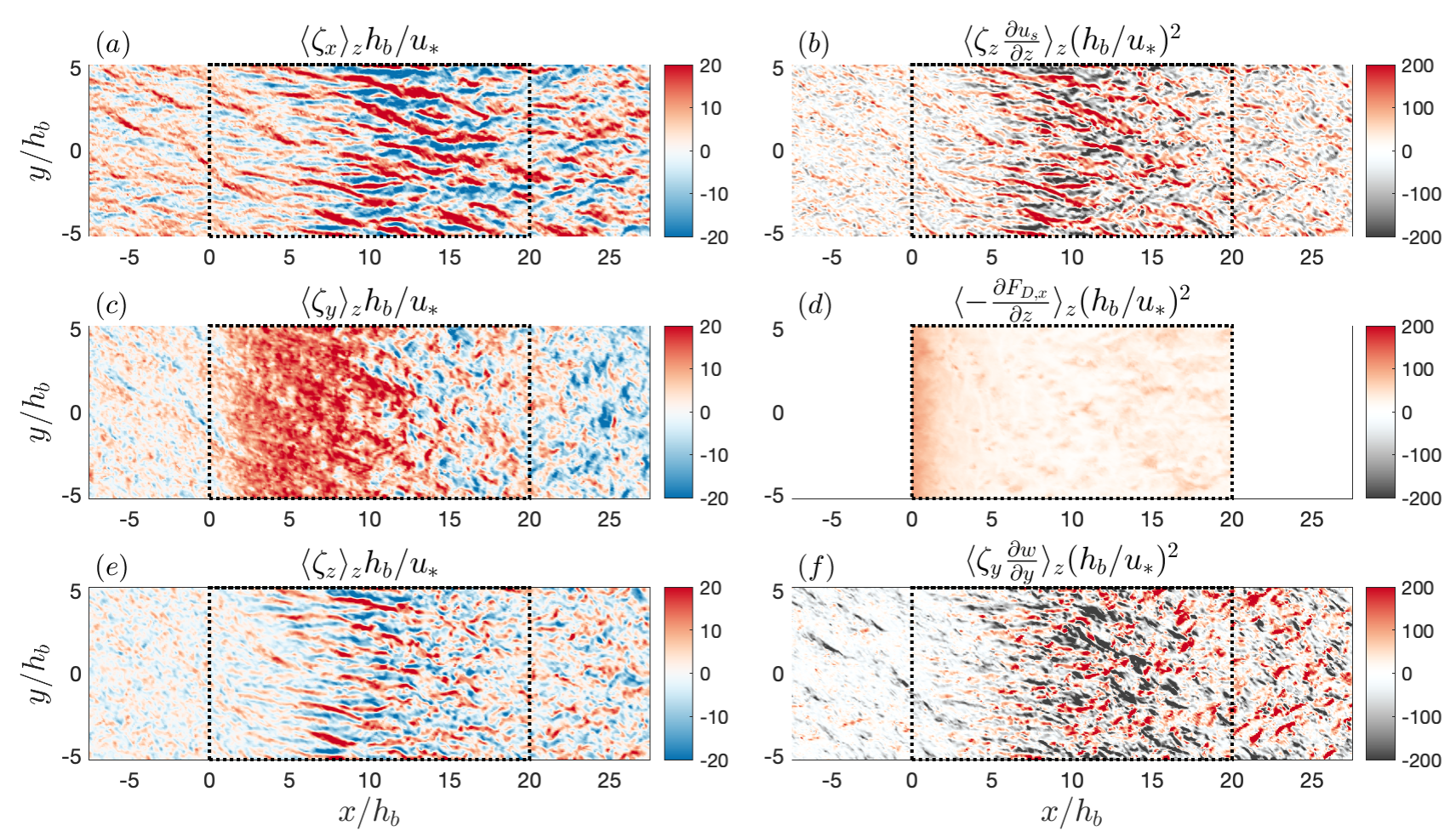}}
  \caption{Top views of vorticity components $\zeta_x$ ($a$), $\zeta_y$ ($c$), 
  $\zeta_z$ ($e$), and their primary forcing terms ($b$, $d$, and $f$) in \eqref{eq:vort-xyz} 
  for case B1H (farm block, harvested profile). 
  The results are snapshots of depth-averaged ($z=0$ to $-h_b$) results. 
  Dotted rectangles show the extent of the farm.}
\label{fig:map-vort-1}
\end{figure}

Subsequently, we investigate the vorticity dynamics in case B1H  
(farm block, harvested profile). Here we analyze a representative 
snapshot as opposed to the time-averaged field, because the 
Langmuir circulation patterns in the farm block are transient 
and will be smeared out through time-averaging. 
The strongest vorticity is generally aligned with the streamwise 
direction, i.e., the $\zeta_x$ component (figure~\ref{fig:map-vort-1}$a$). 
This is particularly evident in regions of intensified Langmuir 
circulation, e.g., around $x/h_b$ from 10 to 15 within the farm 
(comparing figure~\ref{fig:map-vort-1}$a$ with 
figure~\ref{fig:map-w}$b$). The pronounced $\zeta_x$ component 
is consistent with the feature of standard Langmuir circulation 
\citep{mcwilliams1997}. 
The generation of $\zeta_x$ is primarily due to the vorticity 
tilting by the Stokes drift, i.e., $\zeta_z (\partial u_s/\partial z)$, 
similar to that in the kelp rows. 

However, in contrast to the vorticity dynamics in the kelp rows, 
the influence of $\partial F_{D,x}/\partial y$ is diminished in the 
farm block due to the absence of lateral variations in kelp frond density. 
Instead, the vorticity generation is initiated from the leading edge 
of the farm block, where $\partial F_{D,x}/\partial z$, resulting from 
the vertical frond density variability, produces cross-stream vorticity 
$\zeta_y$. The drag force is typically smaller near the surface for the 
harvested profile, which thus leads to a positive vertical shear with 
higher streamwise velocity on top of lower velocity, i.e., positive 
$\zeta_y$ (figure~\ref{fig:map-vort-1}$c$). 
The higher streamwise velocity near the surface tends to 
concentrate toward regions with surface-convergence (indicative of 
downwelling), as a result of lateral and vertical advection by 
Langmuir circulation. Similarly, lower streamwise velocity is 
transported to the upwelling regions in the cross-section. 
The associated lateral variability in streamwise velocity enhances 
vertical vorticity $\zeta_z$ (figure~\ref{fig:map-vort-1}$e$), which 
can subsequently be tilted into $\zeta_x$ by the Stokes drift 
(figure~\ref{fig:map-vort-1}$a$). The increased $\zeta_x$ represents 
stronger Langmuir circulation that can furthur interact with the 
other vorticity components $\zeta_y$ and $\zeta_z$. 
This interaction leads to positive feedback, thereby intensifying 
Langmuir circulation. 

The processes described above are consistent with the CL2 mechanism 
\citep{craik1977,leibovich1977,leibovich1983}, which is widely 
recognized as a key driving factor for standard Langmuir circulation. 
According to the CL2 mechanism, the vorticity tilting due to Stokes 
drift can cause instability and produce Langmuir circulation cells 
when the vertical shear of the Eulerian wind-driven current is in the same direction 
as the vertically sheared Stokes drift. In our study, the presence of the 
kelp farm further enhances the vertical shear in the OML through the 
vertical variability of canopy drag, thus promoting the growth of 
instability and leading to more intensified Langmuir circulation. 

Note that the patterns of $\zeta_y$ and $\zeta_z$ are generally 
consistent with the forcing terms $\partial F_{D,x}/\partial z$ 
and $\zeta_y (\partial w/\partial y)$ in the upstream region of 
the farm, where these vorticity components are initially generated 
(figure~\ref{fig:map-vort-1}). Other vorticity tilting terms become 
more prominent as Langmuir circulation is rapidly enhanced downstream. 
This leads to the more complex vorticity patterns found in the 
downstream part of the farm. 

\begin{figure}
  \centerline{\includegraphics[width=1.1\textwidth]{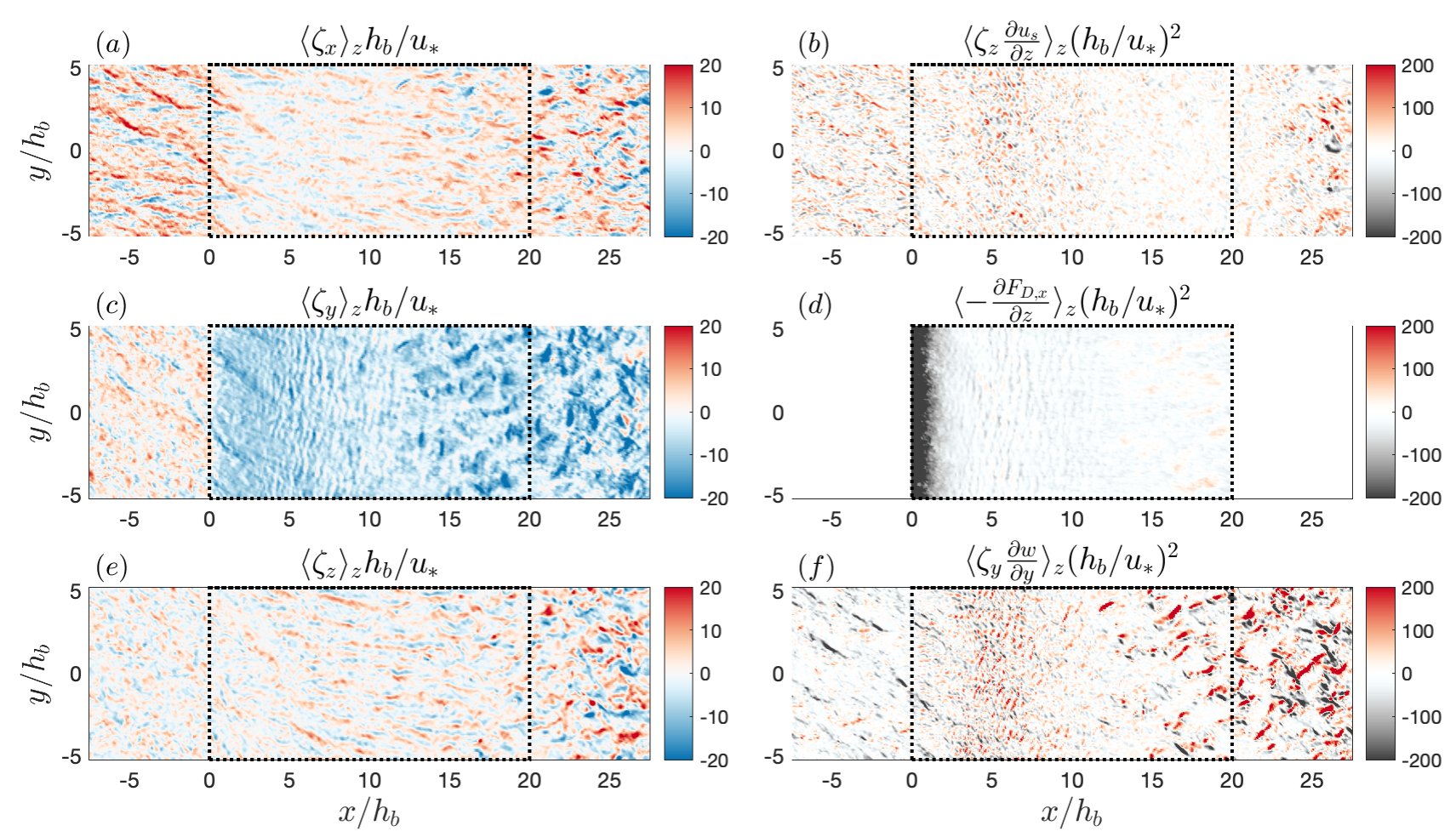}}
  \caption{Top views of vorticity components $\zeta_x$ ($a$), $\zeta_y$ ($c$), 
  $\zeta_z$ ($e$), and forcing terms ($b$, $d$, and $f$) in \eqref{eq:vort-xyz} 
  for case B1R (farm block, ripe profile). 
  These are snapshots of depth-averaged ($z=0$ to $-h_b$) results. 
  Dotted rectangles show the extent of the farm.}
\label{fig:map-vort-2}
\end{figure}

In contrast, in the farm block with the ripe profile (case B1R), 
the presence of the dense layer near the surface tends to decelerate 
the streamwise current. Consequently, negative vertical shear arises 
with lower streamwise velocity on top of higher velocity, in particular 
near the sea surface where the Stokes drift shear is most pronounced 
(figure~\ref{fig:mean-vel-prof}$a$). This leads to the negative 
depth-averaged $\zeta_y$ in figure~\ref{fig:map-vort-2}$(c)$, in contrast 
to the positive  $\zeta_y$ in figure~\ref{fig:map-vort-1}$(c)$ associated 
with the harvested profile (case B1H). The negative vertical shear 
due to the ripe profile in case B1R results in a 
scenario that suppresses the growth of CL2 instability. It thus 
explains the weak vorticity and absence of Langmuir circulation 
patterns in the farm block with the ripe profile 
(figure~\ref{fig:map-vort-2} and figure~\ref{fig:map-w}$c$).

Furthermore, we examine the streamwise evolution of different 
vorticity components. In the farm block with the ripe profile 
(case B1R), cross-stream vorticity $\zeta_y$ promptly increases 
at the leading edge driven by the canopy drag gradient $\partial F_{D,x}/\partial z$. 
Downstream of the leading edge, all the vorticity components 
tend to diminish in magnitude through the farm due to the lack 
of Langmuir circulation generation. In the farm block with the 
harvested profile (case B1H), the rapid growth of $\zeta_y$ at the 
leading edge subsequently causes the increase of $\zeta_z$ 
and $\zeta_x$, as a result of the tilting mechanism. 
In comparison, within the spaced kelp rows (case S26H), the vertical 
vorticity $\zeta_z$ is the vorticity component that initially increases, 
driven by $\partial F_{D,x}/\partial y$ at the lateral edges of kelp 
rows, and this increase in $\zeta_z$ is followed by  
the intensification of $\zeta_x$ and $\zeta_y$. 
This contrast again demonstrates the distinction in vorticity 
generation mechanisms between the two farm configurations. 
It also elucidates why Langmuir turbulence still occurs in kelp rows 
aligned with the current even with the ripe profile (e.g., case S26R): 
$\zeta_z$ can be persistently generated by the spaced kelp rows 
and then tilted by the Stokes drift to produce $\zeta_x$, even if a vertical 
shear that favors the CL2 mechanism is not created. 

\begin{figure}
  \centerline{\includegraphics[width=0.55\textwidth]{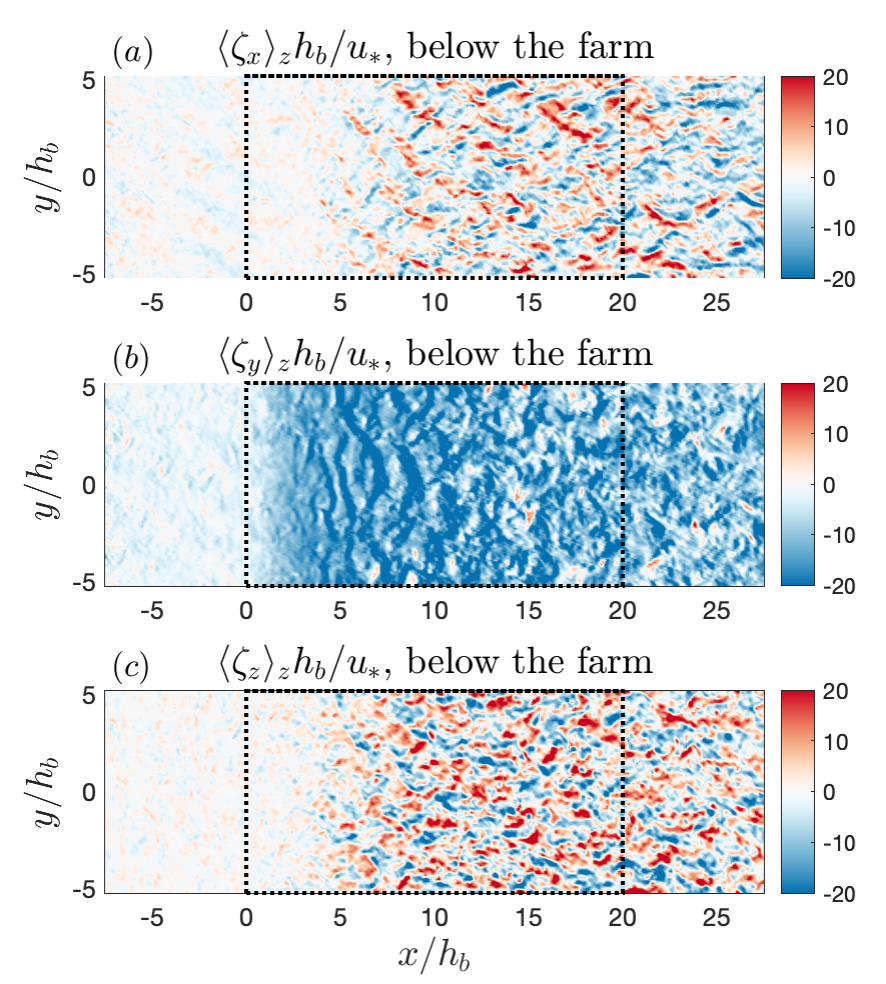}}
  \caption{Top views of vorticity components $\zeta_x$ ($a$), $\zeta_y$ ($b$), 
  and $\zeta_z$  ($c$) for case B1H (farm block, harvested profile). 
  These are snapshots of depth-averaged ($z=-h_b$ to $-2h_b$) results below 
  the farm. 
  Dotted rectangles show the extent of the farm.}
\label{fig:map-vort-1-below}
\end{figure}

Differing from the Langmuir patterns found within the farm, the 
vorticity patterns beneath the farm generally resemble the 
classical shear layer turbulence \citep{finnigan2009,bailey2016} 
(figure~\ref{fig:map-vort-1-below}). Negative cross-stream vorticity 
$\zeta_y$ is generated due to the drag discontinuity 
($\partial F_{D,x}/\partial z$) at the canopy bottom edge. 
Other vorticity components $\zeta_x$ and $\zeta_z$ subsequently 
increase as a result of vorticity tilting. The influence of 
Stokes drift rapidly diminishes with depth, and the 
Stokes drift - vorticity tilting term is thus negligible below the farm. 

\section{Sensitivity to other farm parameters}
\label{sec:other-factors}
\subsection{Effective density}
\begin{figure}
  \centerline{\includegraphics[width=0.8\textwidth]{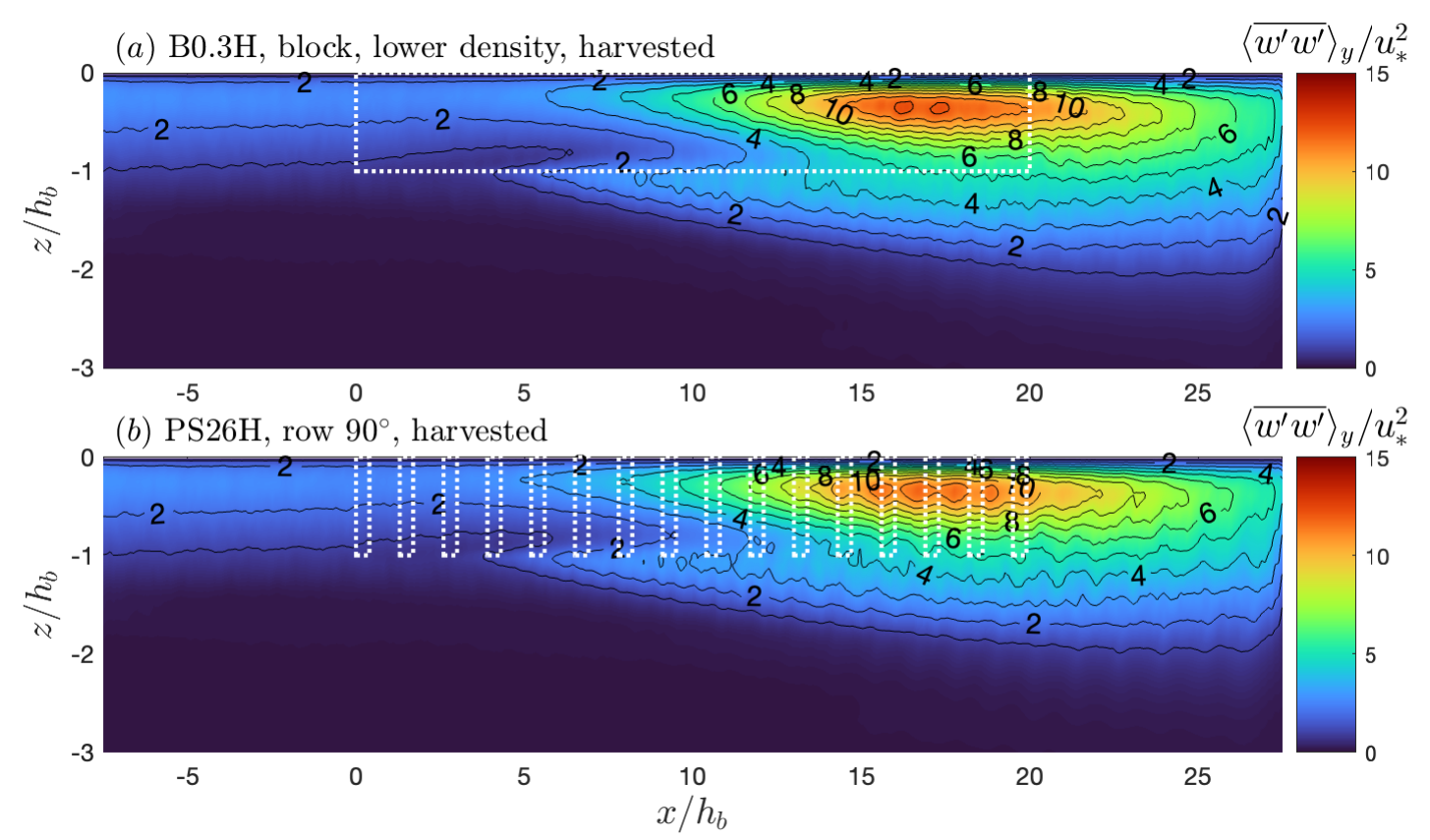}}
  \caption{Side views of the transient component of vertical velocity 
  variance $\left\langle\overline{w'w'}\right\rangle_{y}/u_*^2$. 
  $(a)$:~Case B0.3H, farm block with a lower frond area density, 
  harvested profile. 
  $(b)$:~Case PS26H, spaced rows oriented perpendicular to the 
  geostrophic current, harvested profile.
  The results are temporally and laterally averaged, similar to figure~\ref{fig:sideview-wp2}. 
  Dotted rectangles show the extent of the farm block or rows.
  Note that cases B0.3H and PS26H have an equal effective density 
  $\left\langle a\right\rangle_{xyz}$.}
\label{fig:sideview-wp2-less-dense}
\end{figure}

In the previous sections, we focused on distinctions between 
farm blocks and spaced rows and between different vertical frond 
density profiles. In addition, the farm effective density is also 
an key characteristic affecting the hydrodynamics of suspended 
farms \citep{poggi2004,belcher2012,bailey2013}. 
Figure~\ref{fig:sideview-wp2-less-dense}$a$ shows the turbulence 
kinetic energy in the farm block with a lower effective density 
$\left\langle a\right\rangle_{xyz}$
(case B0.3H, harvested profile, with a density lower than case 
B1H by a factor of 0.3). Overall, the shear layer turbulence 
below the canopy is weaker in case B0.3H compared to B1H, 
consistent with the expected dependence on effective 
density for classical canopy flow \citep{poggi2004,bailey2013}. 

Turbulence intensity within the canopy typically decreases with the 
increased effective density $\left\langle a\right\rangle_{xyz}$ for 
classical canopy flow without waves. This is because a higher 
effective density reduces the penetration of shear layer turbulence 
into the canopy and also enhances the dissipation of turbulence 
by canopy drag. However, in our study turbulence is weaker in 
the lower-density farm block compared to the higher-density case 
(figure~\ref{fig:sideview-wp2-less-dense}$a$ and 
figure~\ref{fig:sideview-wp2}$b$). This is because the lower 
density leads to weaker shear (or vorticity) in the canopy, and 
thus a weaker Stokes drift - vorticity tilting mechanism and 
less Langmuir turbulence production 
(§§~\ref{sec:tke-budget} and \ref{sec:vort-dynamics}). 

Moreover, the region with maximum Langmuir turbulence intensity occurs 
at a larger downstream distance from the farm leading edge in the 
lower effective density case compared to the higher density case. 
This is because it takes a longer distance for flow to adjust to 
canopy drag and for shear to develop with a lower effective density. 
This longer development distance is consistent with 
\eqref{eq:drag-length}, where the adjustment length is expected to 
be inversely proportional to the effective density. 

\subsection{Farm orientation}
The dependence of turbulence generation on farm orientation is also 
examined through a set of simulations with kelp rows oriented 
perpendicular to the direction of current and waves (cases `PS'). 
This is particularly relevant to the farm design problem in realistic 
ocean environments, given the variability of submesoscale processes. 
Farms with kelp rows perpendicular to the flow generally behave 
similarly to farm blocks that have an identical effective density 
$\left\langle a\right\rangle_{xyz}$, e.g., case PS26H 
in figure~\ref{fig:sideview-wp2-less-dense}$(b)$ and case B0.3H in 
figure~\ref{fig:sideview-wp2-less-dense}$(a)$, both of which have a 
same effective density of 0.35~m$^{-1}$. 
Note that the configuration with spaced rows could lead to additional 
stationary streamwise deviations in canopy flow, typically corresponding to 
a dispersive flux \citep{bailey2013,li2019}. Nevertheless, the 
influences of dispersive processes appear to be minimal in the 
examined cases, and the statistics of kelp rows perpendicular to the 
flow are broadly similar to uniform farm blocks (figure~\ref{fig:comp-tke}). 

\subsection{Synthesis}
We have presented cases B0.3H and PS26H 
(refer to table~\ref{tab:para}) above as a representative 
to illustrate the impacts of effective density and farm orientation. 
Other simulations with various density or farm configurations 
generally show a consistent dependence on effective density 
$\left\langle a\right\rangle_{xyz}$. As has been summarized in 
figure~\ref{fig:comp-tke}$(d)$, the intensity of shear layer 
turbulence below the farm positively depends on the effective 
density for all the simulations. In addition, for cases with the 
harvested profile, which favors the generation of Langmuir-type 
turbulence, turbulence intensity within the farm increases as the 
effective density is increased (figure~\ref{fig:comp-tke}$a$). 

By contrast, for the ripe profile, no obvious dependence of 
within-farm turbulence intensity on the effective density is 
found for farm blocks and kelp rows oriented perpendicular to the 
flow (figure~\ref{fig:comp-tke}$a$), because of the absence of 
Langmuir circulation. For kelp rows aligned with the geostrophic flow 
with the ripe profile, turbulence intensity asymtotes toward a 
farm block scenario or no farm scenario with either increasing 
or decreasing effective density, and the maximum intensity is 
achieved with an effective density of around 0.5~m$^{-1}$. 

The distance for farm-generated Langmuir turbulence to peak 
generally increases with the decreased effective density 
$\left\langle a\right\rangle_{xyz}$, for configurations with a farm 
block or kelp rows perpendicular to flow. Nevertheless, in kelp rows 
aligned with flow, the distance for Langmuir turbulence to develop 
is much smaller than a farm block with an identical effective 
density, e.g., comparing figure~\ref{fig:sideview-wp2}$(a)$ with 
figure~\ref{fig:sideview-wp2-less-dense}$(a)$. 
This is because kelp rows aligned with the current can lead to 
rapid growth of Langmuir turbulence through the creation of 
vorticity at lateral edges of rows by the drag force gradient 
$\partial F_{D,x}/\partial y$. It further implies that this type 
of configuration may result in a shorter adjustment region length 
compared to that expected with the effective density in 
\eqref{eq:drag-length}. 

\subsection{Extended farm length}
\begin{figure}
  \centerline{\includegraphics[width=1\textwidth]{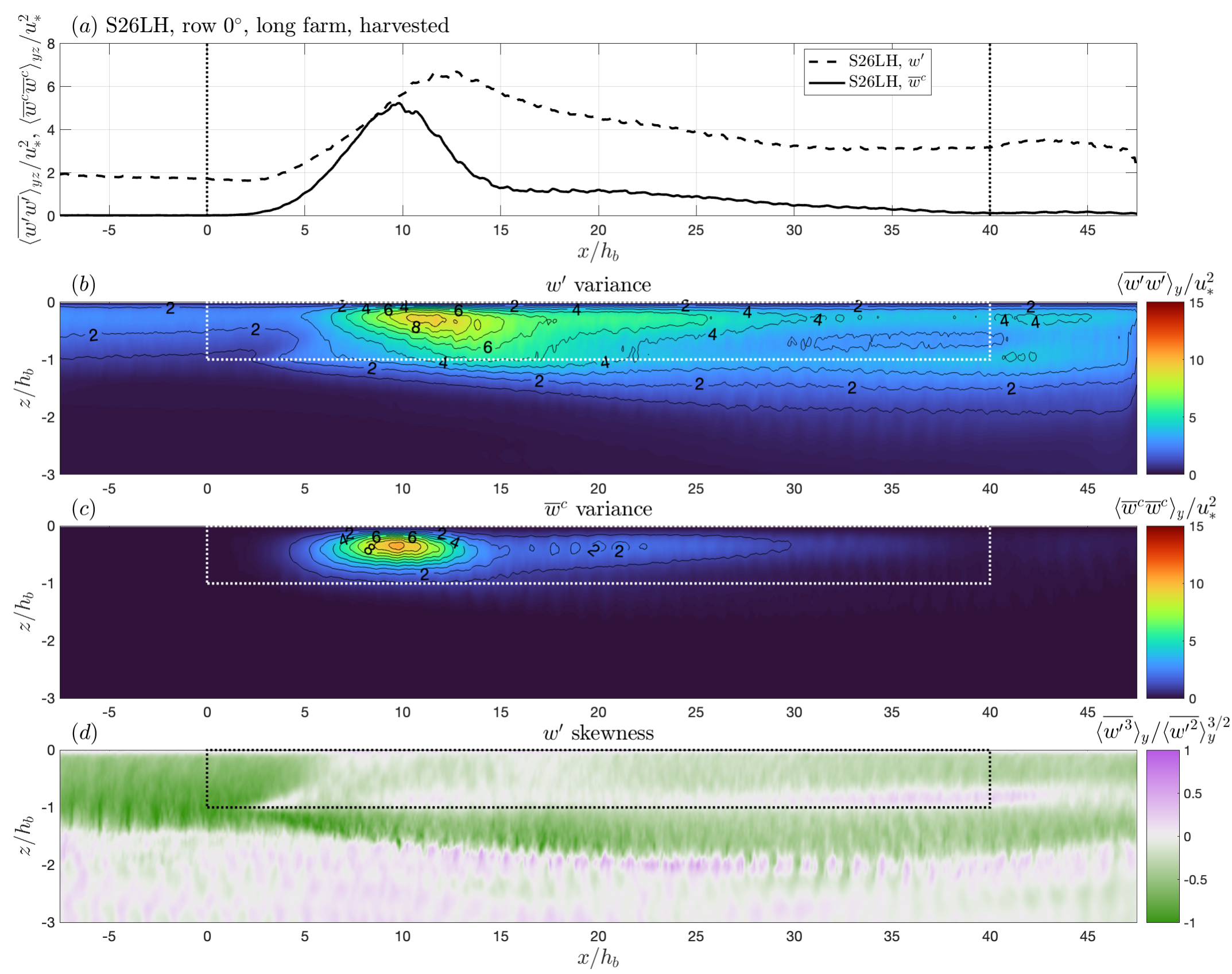}}
  \caption{Turbulence and secondary flow statistics in the long farm 
  simulation (case S26LH, spaced rows aligned with the current, 
  harvested profile). $(a)$: Streamwise variations of vertical velocity variance 
  $\left\langle\overline{w'w'}\right\rangle_{y}/u_*^2$ 
  (turbulence component, dashed line)
  and $\left\langle\overline{w}^c \overline{w}^c\right\rangle_{y}/u_*^2$ 
  (secondary flow component, solid line). 
  The results are temporally, laterally, and vertically ($z=0$ to $-h_b$) averaged. 
  Vertical dotted lines show the extent of the farm. 
  $(b)$ and $(c)$: Side views of $\left\langle\overline{w'w'}\right\rangle_{y}/u_*^2$ 
  and $\left\langle\overline{w}^c \overline{w}^c\right\rangle_{y}/u_*^2$. 
  The results are temporally and laterally averaged. 
  Dotted rectangles show the extent of the farm. 
  $(d)$: Side view of the skewness of $w'$.}
\label{fig:sideview-wp2-long-farm}
\end{figure}

We ran an additional simulation 
(case S26LH, refer to table~\ref{tab:para}) with an extended farm 
length of $L_{MF}=800$~m to further investigate the downstream 
development of farm-generated turbulence. The long farm case S26LH 
has the same configuration as case S26H (spaced kelp rows aligned 
with the flow with $S_{MF}=26$~m, harvested profile). 
The turbulence and secondary flow characteristics in the upstream 
part of the long farm are generally similar to that of the short 
farm, e.g., comparing figure~\ref{fig:sideview-wp2-long-farm} with 
figures~\ref{fig:sideview-wp2}$(a)$, \ref{fig:sideview-wc2}, and 
\ref{fig:sideview-skewness}$(a)$. 
 
In the upstream part of the long farm, secondary flow diminishes 
after around $x/h_b=15$, and turbulence intensity also remains at 
a stable value after a decay from its peak value 
(figure~\ref{fig:sideview-wp2-long-farm}$a$ and $b$). 
Turbulence statistics near the trailing edge of the long farm are 
broadly similar to those of another simulation with an infinite farm 
length (not shown). The canopy flow is thus considered as having 
reached an equilibrium state toward the long farm trailing edge. 

We note that the farm-generated Langmuir-type turbulence and secondary 
flow investigated in this study primarily occurs in the upstream part 
of the farm where flow is adapting to the canopy drag, before the 
establishment of an equilibrium canopy flow. 
Nevertheless, our investigation is relevant to the farm design problem, 
given that the farm length is comparable to or less than the canopy 
adjustment length in a range of realistic farm design considerations. 

The turbulence intensity near the long farm trailing edge, in spite of 
the disappearance of secondary flow there, is still higher than that 
of the standard Langmuir turbulence observed upstream of the farm 
(figure~\ref{fig:sideview-wp2-long-farm}$a$). Moreover, it is 
worthwhile noting that positive skewness of $w'$ only occurs in the 
lower one third of the canopy in the equilibrium state at the trailing 
edge (figure~\ref{fig:sideview-wp2-long-farm}$d$). The positive 
skewness of $w'$ is a typical feature of the shear layer turbulence 
that is penetrated into the canopy, corresponding to the prevalent 
presence of sweep events \citep{raupach1981,katul1997,poggi2004}. 
The negative skewness of $w'$ above the penetrated shear layer is 
indicative of the Langmuir-type turbulence \citep{mcwilliams1997}. 
This indicates that the shear layer turbulence growing from the 
farm bottom does not penetrate through the entire canopy in the 
equilibrium state. 

We calculate the penetration length scale $\delta_e$ based on the 
empirical relationship \citep{nepf2007,nepf2008} 
\begin{linenomath*}
\begin{equation}
\label{eq:pene-length}
    \delta_e = \frac{0.23}{\frac{1}{2}C_D\left\langle a\right\rangle_{xyz}}. 
\end{equation}
\end{linenomath*}
Here the factor of $\frac{1}{2}$ accounts for the projection of frond 
surface area (term $\boldsymbol{P}$ in \eqref{eq:drag}). 
The calculated $\delta_e$ is four times larger 
than the canopy height, suggesting that shear layer eddies would 
penetrate to the sea surface in fully developed canopy flow. 
As a comparison, in the simulation without waves (case S26H-NW), 
positive skewness of $w'$, which represents shear layer turbulence, 
occupies the entire farm height after around $x/h_b=15$, 
and this is consistent with the prediction from \eqref{eq:pene-length}. 
Nevertheless, the coexistence of both positive and negative 
skewness of $w'$ in figure~\ref{fig:sideview-wp2-long-farm}$(d)$ 
implies that wave effects can impede the penetration of shear 
layer eddies. An adjusted OML is established in the equilibrium 
state, characterized by a combination of Langmuir-type turbulence 
and penetrated shear layer turbulence. 

\section{Conclusion}
\label{sec:conclusion}
In this study, we investigate the influence of various suspended 
farm configurations on the OML hydrodynamics. The drag force 
induced by kelp can alter the vertical profile of mean flow and 
leads to generation of shear layer turbulence beneath the farm. 
Moreover, Langmuir circulation can be generated within the farm, 
differing from the standard Langmuir circulation in the absence 
of a farm. These modifications to the OML depend on both the farm 
horizontal arrangement and the vertical profile of frond area density. 

In farm blocks or spaced kelp rows that are perpendicular to the 
current, distinctions in Langmuir circulation patterns emerge 
between the harvested profile and ripe profile (without and with 
the dense layer near the surface, respectively). 
Enhanced Langmuir circulation occurs in the farm with the harvested 
profile, characterized by transient patterns that have a larger 
magnitude compared to the standard Langmuir circulation upstream of 
the farm. The average turbulence intensity in the farm, dominantly 
contributed by Langmuir turbulence, has a positive dependence on 
the effective density $\left\langle a\right\rangle_{xyz}$. 
The generation of Langmuir turbulence is attributed to the 
establishment of a vertically sheared current profile that favors 
CL2 instability, as a result of the vertical variability in canopy 
drag. The lateral vorticity (vertical shear of streamwise velocity) 
gives rise to vertical vorticity, which is subsequently tilted into 
streamwise vorticity by the Stokes drift to create Langmuir circulation. 
By contrast, for the ripe profile, a reversed vertical shear occurs 
near the sea surface due to the presence of the dense kelp layer, 
and Langmuir turbulence is absent in the farm due to the 
suppression of CL2 instability. 

The farm configuration involving spaced kelp rows aligned with the 
flow leads to the generation of attached Langmuir circulation for both 
the harvested and ripe profiles. The attached Langmuir circulation 
is spatially locked in the lateral direction, comprised of both a 
turbulence component and a stationary secondary flow component. 
Strong lateral shear is generated at the lateral edges of kelp rows, 
and this vertical vorticity is subsequently tilted into the streamwise 
direction by the Stokes drift, producing attached Langmuir circulation. 
The intensity of Langmuir turbulence monotonically increases with 
the effective density $\left\langle a\right\rangle_{xyz}$ for the 
harvested profile. However, for the ripe profile, the enhancement 
of Langmuir turbulence diminishes at either high $\left\langle a \right\rangle_{xyz}$, 
resembling a farm block, or low $\left\langle a \right\rangle_{xyz}$, 
resembling a scenario without a farm. The peak value occurs at an 
intermediate effective density, which is around 
$\left\langle a\right\rangle_{xyz}=0.5$m$^{-1}$ in our simulations. 
Additionally, the strength of the secondary flow increases from the 
kelp row lateral spacing $S_{MF}$ of 13~m to 52~m, and decays at 
$S_{MF}=208$~m. The simulations do not allow for an accurate determination 
of the maximum secondary flow intensity with respect to $S_{MF}$, 
and it appears to fall within the range of $S_{MF}=52$~m and 208~m. 

The vertical shear of mean flow is enhanced below the farm, leading to 
development of shear layer turbulence. The intensity of shear layer 
turbulence generally increases with the effective density 
$\left\langle a\right\rangle_{xyz}$, while displaying a minimal 
dependence on the specific farm configuration. 

We have focused on a specific set of representative oceanic conditions 
in the present study, and the implications of turbulence generation 
can be extended to various wave and current conditions. 
Stronger winds and waves typically increase the Stokes drift 
and the CL vortex force, therefore further intensifying Langmuir 
turbulence within the farm configurations that favor turbulence generation. 
Weaker ocean currents would lead to a reduced level of shear layer turbulence 
below the farm, as its generation depends on the magnitude of shear 
(figure~\ref{fig:comp-tke}$d$ and $e$). Moreover, the enhancement of 
Langmuir turbulence in the farm relies on the interaction between the 
Stokes drift and farm-modulated currents. The canopy drag effect would 
diminish under conditions of weaker ocean currents, which is thus 
expected to result in less deviation from the standard Langmuir turbulence. 

The different mean flow, secondary flow, and turbulence characteristics 
associated with various farm configurations have significant implications 
for nutrient transport and farm performance, as will be examined in 
an upcoming companion study. 
Moreover, the presence of suspended farms notably alters turbulence 
intensity within the OML. 
These findings underscore the potential of floating obstacle structures 
in reshaping characteristics of the ocean surface boundary layer. 
The presence of such floating structures may enhance 
the shear in ocean currents, giving rise to shear layer turbulence. 
Additionally, the modified velocity profile caused by floating structures 
can interact with surface gravity waves, thereby influencing the 
generation of Langmuir turbulence in the OML. 

\vspace*{\baselineskip}

\backsection[Acknowledgements]{The authors thank Kristen Davis 
for helpful discussions.}

\backsection[Funding]{The research leading to these results was funded by 
the ARPA-E MARINER Program DE-AR0000920. We would like to acknowledge 
high-performance computing support from Cheyenne (doi:10.5065/D6RX99HX) 
provided by NCAR's Computational and Information Systems Laboratory, 
sponsored by the National Science Foundation. Chao Yan is supported by the youth innovation team of China Meteorological Administration (No. CMA2024QN12).}

\backsection[Declaration of interests]{The authors report no conflict of interest.}

\appendix
\section{Detailed farm parameters}\label{app-farm-para}
\begin{table}
  \begin{center}
  \begin{tabular}{lccccccc}
      Case  & Wave & $L_{MF}$ (m)  & $S_{MF}$ (m) & $W_{MF}$ (m) & Frond vert. prof. &      
            Orientation & $\left\langle a\right\rangle_{xyz}$ (m$^{-1}$) \\[8pt]
   \rowcolor[gray]{0.9}
       B1H\ $\star$  & Yes & 400 & - & 208 & Harvested & - & 1.14 \\
   \rowcolor[gray]{0.9}
       B0.3H  & Yes & 400 & - & 208 & Harvested & - & 0.35 \\
   \rowcolor[gray]{0.9}
       B0.16H  & Yes & 400 & - & 208 & Harvested & - & 0.18 \\
   \rowcolor[gray]{0.9}
       B0.08H  & Yes & 400 & - & 208 & Harvested & - & 0.09 \\
   \rowcolor[gray]{0.9}
       B1R\ $\star$  & Yes & 400 & - & 208 & Ripe & - & 2.20 \\
   \rowcolor[gray]{0.9}
       B0.3R  & Yes & 400 & - & 208 & Ripe & - & 0.68 \\
   \rowcolor[gray]{0.9}
       B0.16R  & Yes & 400 & - & 208 & Ripe & - & 0.35 \\
   \rowcolor[gray]{0.9}
       B0.08R  & Yes & 400 & - & 208 & Ripe & - & 0.18 \\
       &\\
       S26H\ $\star$  & Yes & 400 & 26 & 8 & Harvested & Aligned & 0.35 \\
       S13H  & Yes & 400 & 12 & 8 & Harvested & Aligned & 0.70 \\
       S16H  & Yes & 400 & 16 & 8 & Harvested & Aligned & 0.57 \\
       S52H  & Yes & 400 & 52 & 8 & Harvested & Aligned & 0.18 \\
       S52W20H  & Yes & 400 & 52 & 20 & Harvested & Aligned & 0.44 \\
       S208H  & Yes & 400 & 208 & 8 & Harvested & Aligned & 0.04 \\
       S26R  & Yes & 400 & 26 & 8 & Ripe & Aligned & 0.67 \\
       S13R  & Yes & 400 & 13 & 8 & Ripe & Aligned & 1.35 \\
       S16R  & Yes & 400 & 16 & 8 & Ripe & Aligned & 1.10 \\
       S52R  & Yes & 400 & 52 & 8 & Ripe & Aligned & 0.34 \\
       S52W20R  & Yes & 400 & 52 & 20 & Ripe & Aligned & 0.85 \\
       S208R  & Yes & 400 & 208 & 8 & Ripe & Aligned & 0.08 \\[5pt]
       S26LH  & Yes & 800 & 26 & 8 & Harvested & Aligned & 0.35 \\
       &\\
   \rowcolor[gray]{0.9}
       PS26H  & Yes & 400 & 26 & 8 & Harvested & Perpendicular & 0.35 \\
   \rowcolor[gray]{0.9}
       PS52H  & Yes & 400 & 52 & 8 & Harvested & Perpendicular & 0.18 \\
   \rowcolor[gray]{0.9}
       PS52W20H  & Yes & 400 & 52 & 20 & Harvested & Perpendicular & 0.44 \\
   \rowcolor[gray]{0.9}
       PS26R  & Yes & 400 & 26 & 8 & Ripe & Perpendicular & 0.67 \\
   \rowcolor[gray]{0.9}
       PS52R  & Yes & 400 & 52 & 8 & Ripe & Perpendicular & 0.34 \\
   \rowcolor[gray]{0.9}
       PS52W20R  & Yes & 400 & 52 & 20 & Ripe & Perpendicular & 0.85 \\
       &\\
       PRE & Yes & - & - & - & - & - & -  \\[8pt]
       B1H-NW & No & 400 & - & 208 & Harvested & - & 1.14 \\
       B1R-NW & No & 400 & - & 208 & Ripe & - & 2.20 \\
       S26H-NW & No & 400 & 26 & 8 & Harvested & Aligned & 0.35 \\ 
       PRE-NW & No & - & - & - & - & - & - \\
  \end{tabular}
  \caption{Farm parameters. The cases selected for detailed 
  analysis are marked with stars. The letters `H' and `R' 
  denote  harvested and ripe profiles, respectively. The letter 
  `B' represents farm block cases, and 1, 0.3, 0.16, and 0.08 
  are the frond density multiplication factors, influencing the 
  effective density $\left\langle a\right\rangle_{xyz}$. The letter 
  `S' represents spaced kelp rows aligned with $x$-direction, 
  and `PS' represents spaced kelp rows oriented 
  perpendicular to $x$-direction. Numeric values 13, 16, 26, 52, 
  and 208 denote the row spacing parameter ($S_{MF}$, in meters). 
  Most configurations have a row width $W_{MF}$ of 8~m, and 
  several cases have a row width of 20~m, denoted by `W20'. 
  Additionally, a long farm simulation is named as S26LH, 
  characterized by an extended farm length $L_{MF}$ of 800~m. 
  The precursor simulation in absence of a farm is refered to as PRE. 
  The term `-NW' is used to denote the simulations conducted without 
  any surface wave forcing, meaning the Stokes drift velocity is zero.}
  \label{tab:para}
  \end{center}
\end{table}

A range of farm simulations with spaced rows are conducted (table~\ref{tab:para}), 
where the spacing parameter $S_{MF}$ varies from 13~m to 208~m. 
The kelp row width is $W_{MF}=8$~m for different $S_{MF}$. 
There is also an additional case with an increased row width 
of $W_{MF}=20$~m for $S_{MF}=52$~m. We primarily focus on cases 
with the farm length of $L_{MF}=400$~m, and a long farm 
simulation with $L_{MF}=800$~m is also investigated for 
$S_{MF}=26$~m and $W_{MF}=8$~m. 

For farm block simulations, each frond density profile is 
additionally multiplied by a factor of 0.3, 0.16, and 0.08, 
to investigate the influence of decreased 
effective density (table~\ref{tab:para}). Note that the kelp farm block simulations with 
a multiplication factor of 0.3 and 0.16 result in an identical 
effective density $\left\langle a \right\rangle_{xyz}$ to the 
spaced kelp row simulations with $S_{MF}=26$~m and $S_{MF}=52$~m, 
respectively, for a row width of $W_{MF}=8$~m. 

\section{A single kelp row}\label{app-single-row}
\begin{figure}
\centerline{\includegraphics[width=0.7\textwidth]{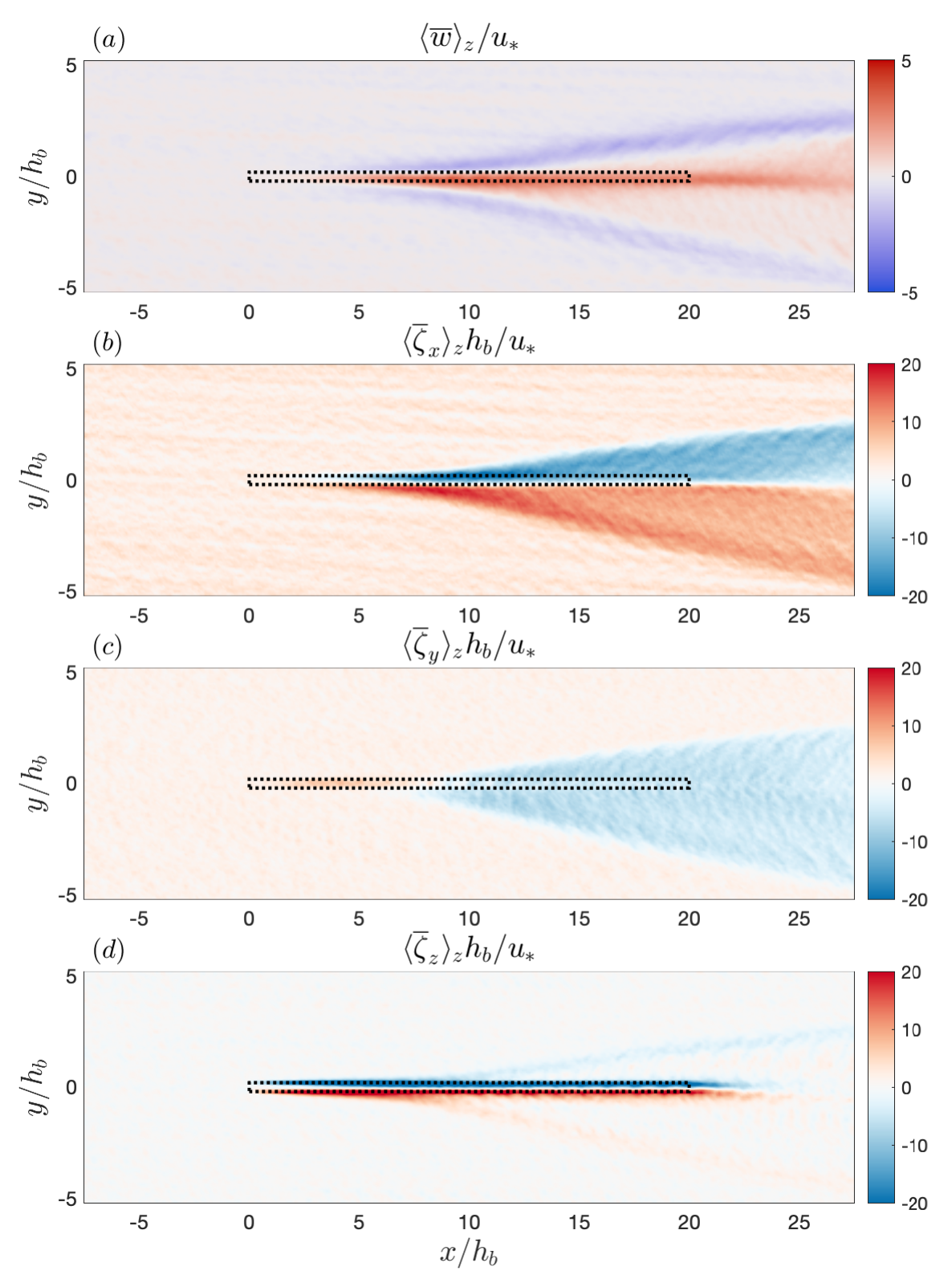}}
  \caption{Top views of vertical velocity $w$ $(a)$ and vorticity components 
  $\zeta_x$, $\zeta_y$, and $\zeta_z$ (b-d) for case S208H 
  (a single row aligned with the current, harvested profile). 
  These results are time-averaged and depth-averaged ($z=0$ to $-h_b$). 
  Dotted rectangles show the extent of the kelp row.}
\label{fig:map-vort-single-row}
\end{figure}

Figure~\ref{fig:map-vort-single-row} shows the simulation of a single 
kelp row aligned with the current and waves (case S208H, harvested 
profile, corresponding to a lateral spacing $S_{MF}$ of 208~m under 
periodic boundary conditions in $y$-direction). 
The regions with strong farm-generated vertical vorticity $\zeta_z$ 
remain close to the lateral edges of the kelp row. The 
vertical velocity $w$ and streamwise vorticity $\zeta_x$, which 
correspond to the attached Langmuir circulation, exhibit a 
downstream expansion in width. 
This lateral expansion of Langmuir circulation, i.e., the Langmuir 
wake pattern associated with the farm, can thus influence the 
horizontal distribution of farm-generated turbulence and secondary flow. 
Moreover, in kelp rows with smaller lateral spacing $S_{MF}$,  
the expansion of the wake may encounter interference due to the 
presence of other neighboring rows.

\bibliographystyle{jfm}
\bibliography{references}

\end{document}